\documentclass{aa}  

\newcommand{\halofdca }{\textsc{Halo-FDCA}}
\newcommand{\halofdcaE }{Halo-Flux Density CAlculator}

\newcommand{\xmm }{{\em XMM-Newton}}
\newcommand{\chandra }{{\em Chandra}}

\newcommand{\planck }{{\em Planck}}

\newcommand{\lofar }{LOFAR}
\newcommand{\lofarE }{LOw Frequency ARray}

\newcommand{\lotss }{LoTSS}
\newcommand{\lotssE }{LOFAR Two-meter Sky Survey}
\newcommand{\lolss }{LoLSS}
\newcommand{\lolssE }{LOFAR LBA Sky Survey}

\usepackage{graphicx}
\usepackage{amssymb}
\usepackage{multirow,bigdelim}
\usepackage{txfonts}
\usepackage[export]{adjustbox}
\usepackage{color}
\usepackage[normalem]{ulem}
\usepackage{xcolor}

\begin{document} 

\title{The Planck clusters in the LOFAR sky}
\subtitle{V. LoTSS-DR2: Mass - radio halo power correlation at low frequency}

\authorrunning{V. Cuciti et al.} 
\titlerunning{The Planck clusters in the LOFAR sky}

\author{V. Cuciti\inst{\ref{unibo},\ref{ira},\ref{hamburg}}, R. Cassano\inst{\ref{ira}}, M. Sereno\inst{\ref{oas}, \ref{infn}}, G. Brunetti\inst{\ref{ira}}, A. Botteon\inst{\ref{ira}}, T. W. Shimwell\inst{\ref{astron}, \ref{leiden}}, L. Bruno\inst{\ref{unibo}, \ref{ira}}, F. Gastaldello \inst{\ref{iasf}}, M. Rossetti\inst{\ref{iasf}}, X. Zhang\inst{\ref{mpe}}, A. Simionescu\inst{\ref{sron}, \ref{leiden}, \ref{kavli}}, M. Br{\"u}ggen\inst{\ref{hamburg}}, R. J. van Weeren\inst{\ref{leiden}}, A. Jones\inst{\ref{hamburg}}, H. Akamatsu\inst{\ref{sron}}, A. Bonafede\inst{\ref{unibo}, \ref{ira}}, F. De Gasperin\inst{\ref{ira}, \ref{hamburg}}, G. Di Gennaro\inst{\ref{hamburg}}, T. Pasini\inst{\ref{hamburg}}, H. J. A. R{\"o}ttgering\inst{\ref{leiden}}  }

\institute{
Dipartimento di Fisica e Astronomia, Universit\`{a} di Bologna, via P.~Gobetti 93/2, I-40129 Bologna, Italy \label{unibo}
\email{virginia.cuciti@unibo.it}
\and
INAF - IRA, via P.~Gobetti 101, I-40129 Bologna, Italy \label{ira}
\and
Hamburger Sternwarte, Universit\"{a}t Hamburg, Gojenbergsweg 112, D-21029 Hamburg, Germany \label{hamburg}  
\and
Leiden Observatory, Leiden University, PO Box 9513, 2300 RA Leiden, The Netherlands \label{leiden}
\and
INAF - IASF Milano, via A.~Corti 12, I-20133 Milano, Italy \label{iasf}
\and
ASTRON, the Netherlands Institute for Radio Astronomy, Postbus 2, NL-7990 AA Dwingeloo, The Netherlands \label{astron}
\and
Max Planck Institute for Extraterrestrial Physics, Giessenbachstrasse 1, 85748 Garching, Germany \label{mpe}
\and
SRON Netherlands Institute for Space Research, Niels Bohrweg 4,
2333 CA Leiden, The Netherlands \label{sron}
\and
Kavli Institute for the Physics and Mathematics of the Universe, The University of Tokyo, Kashiwa, Chiba 277-8583, Japan \label{kavli}
\and
INAF - Osservatorio di Astrofisica e Scienza dello Spazio, via P.~Gobetti 93/3, I-40129, Bologna, Italy \label{oas}
\and
INFN, Sezione di Bologna, viale Berti Pichat 6/2, I-40127, Bologna, Italy \label{infn}
}

\date{Received XXX; accepted YYY}

\abstract
{Many galaxy clusters show diffuse cluster--scale emission in the form of radio halos, showing that magnetic fields and relativistic electrons are mixed in with the intra-cluster medium (ICM). There is general agreement that the origin of radio halos is connected to turbulence, generated during cluster mergers.  
Statistical studies of large samples of galaxy clusters in the radio band have the potential to unveil the connection between the properties of radio halos and the mass and dynamics of the host clusters.
}
{Previous studies have been limited to massive clusters and based on a small number of radio halos. The aim of this paper is to investigate the scaling relation between the radio power of radio halos and the mass of the host clusters at low frequencies and down to lower cluster masses.}
{We analysed the clusters from the second catalogue of \planck\ Sunyaev Zel'dovich sources that lie within the 5634 deg$^2$ covered by the second Data Release of the \lotssE. We derived the correlation between the radio power and the mass of the host clusters and we investigated the distribution of clusters without radio halos with respect to the correlation. We use X-ray observations to classify the dynamical state of clusters and investigate its role on the power of radio halos.}
{We found a correlation between the power of radio halos at 150 MHz and the mass of the host clusters down to $3 \times 10^{14}M_\odot$. This correlation has a large scatter, part of which can be attributed to the different dynamical states of host clusters. We used two statistical test to show that the distribution of clusters with and without (upper limits) radio halos in the mass-radio power diagram is not compatible with a single correlation and that it is also not compatible with clusters being uniformly distributed below an upper envelope constituted by the correlation.}
{}

\keywords{radiation mechanisms: non-thermal -- galaxies: clusters: intracluster medium -- galaxies: clusters: general -- radio continuum: general -- X-rays: galaxies: clusters -- acceleration of particles}

\maketitle
%

\section{Introduction}

Mergers between galaxy clusters are the final stages of cosmic structure formation and the most energetic events in the current Universe, releasing up to $\sim10^{63}$ ergs in a Gyr timescale. During these events shocks and turbulence are induced into the intra-cluster medium (ICM) which amplify cluster magnetic fields and accelerate particles to relativistic speeds. Radiation from these non-thermal phenomena in galaxy clusters is observable in the radio band in the form of radio halos and radio relics \citep[e.g.][for an observational review]{vanweeren19}. Radio halos are found at the centres of clusters and their emission roughly follows the X-ray emission from the thermal ICM. Radio relics, on the other hand, are located at the periphery of clusters and have elongated morphologies. In the current theoretical picture, radio halos are powered by turbulence, whereas radio relics trace shock waves propagating through the ICM \citep[][for a theoretical review]{brunettijones14}. Both, radio halos and relics are thus probes of the dissipation of the gravitational energy from large to smaller scales. 

In this paper, we focus on radio halos. Studying the statistical properties of radio halos in galaxy clusters has become increasingly important in the last decade as it is a powerful tool to unveil the connection
and evolution of the non-thermal cluster-scale emission with, both, cluster dynamics and cluster mass. Such studies can test the theoretical models for the origin of radio halos. Previous studies, such as the Giant Metrewave Radio Telescope (GMRT) radio halos survey \citep{venturi07,venturi08} and its extension \citep{kale13, kale15} have revealed a statistical connection between the presence of radio halos and the merging status of the host clusters \citep{cassano10, cuciti15, cuciti21b}, supporting the idea that mergers play a key role in the formation of radio halos. Moreover, a correlation exists between the radio power of halos and the mass of the host clusters \citep{basu12,cassano13, cuciti21b, duchesne21, vanweeren21, george21}. The upper limits on the radio power of clusters with no diffuse emission populate a different region in the mass--radio power diagram, typically being about a factor of four below the correlation \citep{cassano13, cuciti21b}. 

According to turbulent re-acceleration models, the radio spectra of radio halos should show a cutoff at a frequency that is proportional to the energy available in the merger \citep{cassanobrunetti05}. In particular, the mass of clusters is one of the parameters that sets the energy budget for particle acceleration. Therefore, massive systems undergoing major mergers are expected to form radio halos with a typical spectral index of $\alpha\sim-1.3$ (with F$_{\nu}\propto \nu^{\alpha}$) visible up to $\sim$GHz frequencies before rapidly steepening, while less massive systems and minor mergers are expected to generate Ultra-Steep Spectrum radio halos (USSRHs, $\alpha<-1.5$), effectively detectable only at $\sim$100 MHz \citep{cassano12}. However, historically these studies were performed at frequencies of $\sim1$ GHz and this limited the number of detections of radio halos and the mass range that could be explored to $M_{500}>5-6\times 10^{14} M_\odot$ \citep{cuciti21b}. 

In this respect, \lofarE\ (\lofar) has enabled observations of galaxy clusters at frequencies $<$200 MHz with unprecedented sensitivity and resolution. \lofar\ is carrying out sensitive wide area surveys of the entire Northern sky at 120$-$168 MHz and 42$-$66 MHz in the context of the \lotssE\ \citep[\lotss;][]{shimwell17} and \lolssE\ \citep[\lolss;][]{degasperin21}, respectively. One of the main goals of these surveys is the discovery of diffuse Mpc-scale radio sources in galaxy clusters providing large samples suitable to perform statistical studies. Based on turbulent acceleration models, \citet{cassano10a} predicted that the LOFAR survey at 150 MHz should detect $\sim350$ radio halos in the Northern Sky.
A first step in this direction has been carried out by \cite{vanweeren21} who analysed the diffuse emission in 26 galaxy clusters selected from the second \planck\ catalogue of Sunyaev Zel'dovich (SZ) sources \citep[PSZ2;][]{planck16} that lie within the 424 deg$^2$ of the first \lotss\ Data Release \citep[\lotss-DR1;][]{shimwell19}. They found eight radio halos, therefore they needed to complement their measurements with literature information to study the mass--radio power correlation.

Recently, we have started a large project\footnote{Images, tables, and further information of all targets can be found on the project website \url{https://lofar-surveys.org/planck_dr2.html}.} with the aim of studying the statistical properties of the diffuse sources in galaxy clusters at 150 MHz. To this purpose, we selected all the clusters from the second PSZ2 catalogue that have been covered by the second \lotss\ Data Release (\lotss-DR2; Shimwell et al. 2022).
In \citet{botteon22} (Paper I) we present the sample, describe the data analysis, classify the cluster radio sources, and provide the measurements of different quantities. In \citet{bruno23} (Paper II) we present the procedure and derive the upper limits to the radio power of clusters with undetected diffuse emission. In \citet{zhang23} (Paper III) we analysed the available X-ray, \chandra\ and/or \xmm\, archival data to derive the morphological properties and the ICM density perturbations of clusters. In \citet{cassano23} (Paper IV) we study the occurrence of radio halos as a function of mass and redshift, the connection between radio halos and the dynamics of clusters and compare the results with the expectation from theoretical models. In \citet{jones23} (Paper VI) we focus on radio relics and discuss their occurrence and their scaling relations. Here, we present the correlation between the radio power of halos and the masses of the host clusters (Section \ref{sec:correlation}), and discuss the scatter of this correlation and the distribution of clusters with and without radio halos (Section \ref{Sec:scatter} and \ref{sec:bimodality}). Finally, we compare our results to previous work at higher frequencies (Section \ref{sec:comparison}).

\section{The sample}
\label{sec:sample}
\begin{figure}
\centering
\includegraphics[width=\hsize]{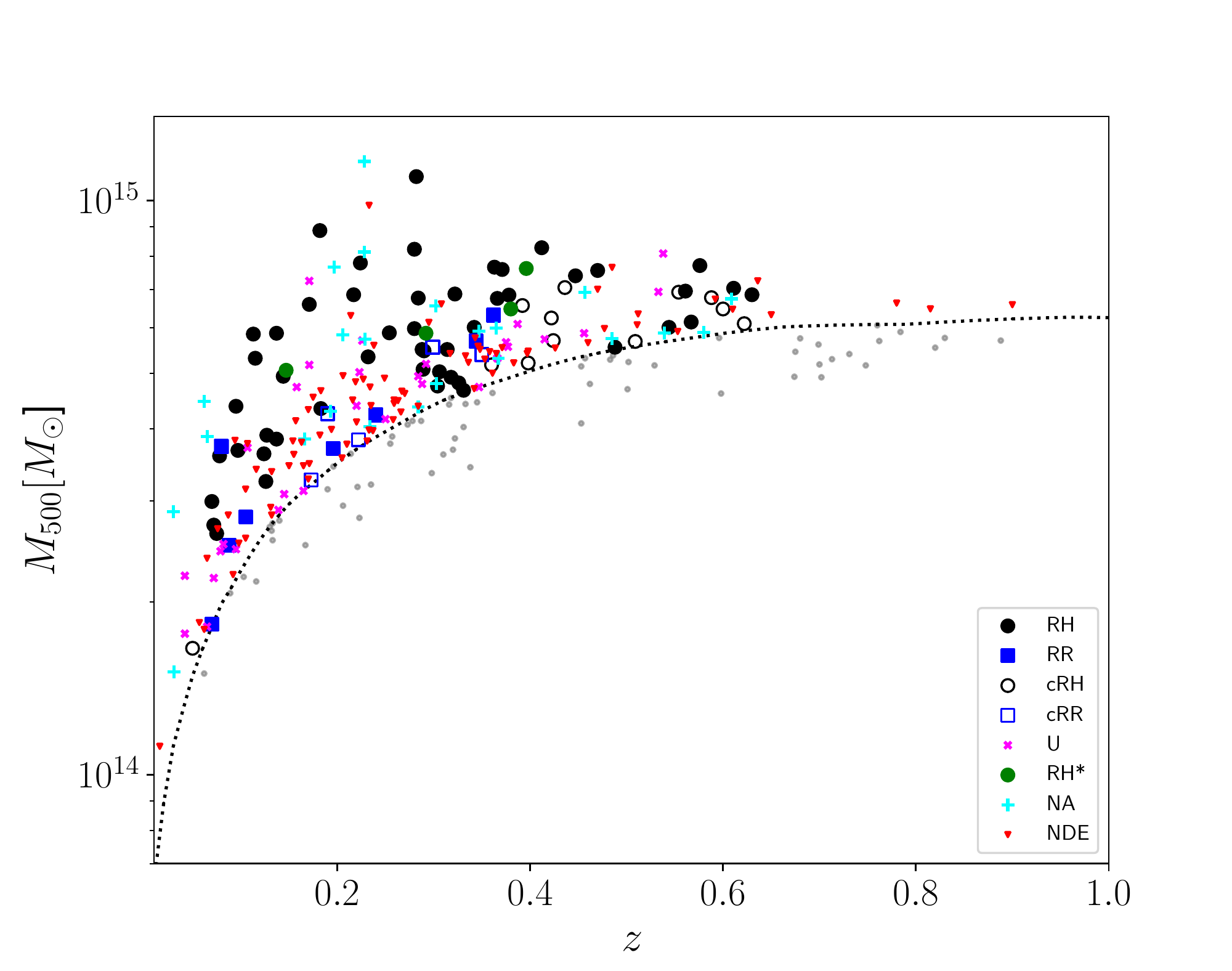}
\caption{Mass--redshift distribution of the clusters of the sample. Clusters above the 50\% {\it Planck} completeness line (dotted line) are marked depending on their classification. Grey points represent clusters below the 50\% completeness line and are not considered in this paper.}
\label{fig:M_z_compl50}
\end{figure}

The PSZ2 catalog \citep{planck16} contains 1653 SZ-sources detected over the entire sky, of which 309 lie in the \lotss-DR2 footprint, and 281 out of 309 have been confirmed as clusters and their mass and redshift have been measured.
In Paper I, we classified the diffuse radio emission in those clusters by visually inspecting a set of \lofar\ images at different resolutions (with and without source subtraction) together with the optical/X-ray overlay images. In order to make this classification objective and reproducible, we used a decision tree that we followed during the inspection of the images to classify the diffuse emission. The radio sources are classified as: \textit{Radio halos} (RH), extended sources that occupy the region where the bulk of the X-ray emission from the ICM is detected; 
\textit{Radio relics} (RR): elongated sources whose position is offset from the bulk of the X-ray emission from the ICM; 
\textit{Candidate radio halos/relics} (cRH/cRR): when the diffuse emission has the appearance of a RH/RR, but the absence of \chandra\ or \xmm\ X-ray observations do not allow to firmly claim this;
\textit{Uncertain} (U): if the emission is either significantly affected by calibration/subtraction artefacts or if it does not fall in the categories of radio halos and relics;
\textit{No diffuse emission} (NDE): these objects do not show the presence of diffuse emission that is not associated with AGN;
\textit{Not applicable} (N/A): when the image quality is not adequate to properly classify the emission.

We used the \halofdcaE\footnote{\url{https://github.com/JortBox/Halo-FDCA}} \citep[\halofdca;][]{boxelaar21} to measure the integrated flux density of radio halos. We adopted an exponential profile to fit the surface brightness of radio halos. This model depends two parameters: the central surface brightness $I_0$ and the \textit{e}-folding radius ($r_e$). As suggested by \citet{murgia09}, when calculating the \halofdca\ derived integrated flux densities we integrated the best-fit models up to a radius of three times the \textit{e}-folding radius. This choice leads to a flux density which is $\sim$80\% of the one that would be obtained by integrating the model up to infinity and is motivated by the fact that halos do not extend indefinitely. The $k$-corrected radio power of the sources at 150 MHz are derived according to the usual formula and assuming a power law radio spectra with spectral index $\alpha=-1.3$ (see Eq.5 in Paper I), which is typical of radio halos \citep[e.g.][]{feretti12, vanweeren19}. We note that some radio halos may have steeper spectra. Hence, the reader should take the values of $P_{150\rm{MHz}}$ with caution, especially for high redshift clusters, where the $k$-correction makes the value of $P_{150\rm{MHz}}$ more affected by the adopted $\alpha$. In a few cases the fitting could not be performed reliably owing to the low signal to noise of the diffuse emission. 
These sources are marked as RH*/cRH* and their integrated flux densities cannot be determined accurately with current data. 

For 75 galaxy clusters, out of 114 NDE with available redshift, we derived the upper limit on the radio power of a possible halo by simulating radio halos through the injection of mock visibilities in the LOFAR uv-data and assuming negligible flux absorption due to calibration. The injection technique and the resulting upper limits are reported in Paper II.

Here we focus on a subsample of 221 clusters that lie above the 50\% {\it Planck} completeness line (Fig.~\ref{fig:M_z_compl50}). This means that we may have missed up to 50\% of clusters above the completeness line. We do not expect this choice to introduce biases in our results since there are no significant differences between the completeness functions for relaxed and disturbed clusters, as shown by \citet{planck16}. This choice represents a reasonable compromise between the size and the completeness of the sample.
Among these 221 clusters three are excluded since they lie within the Coma cluster region \citep[which requires a special data reduction scheme given its large angular extent,][]{bonafede21,bonafede22} and one is PSZ2 G060.10+15.59 for which the ionospheric conditions at the time of the observation did not allow a proper direction-dependent calibration, hence it is excluded.
Moreover, we will not consider the radio halo in PSZ2 G166.62+42.13 because its emission is contaminated by the emission of multiple radio relics and the radio halo power was not computed for this reason \citep{botteon22}. Considering these exclusions, the number of clusters analysed in this paper is 216, we list the number of clusters per category in Table \ref{Tab:numbers}.

\begin{table} 
\begin{scriptsize}
\begin{center}
\caption{Number of clusters per category}
\begin{tabular}{lcccccccc}

&	RH	&	RR	&	cRH	&	cRR  & RH* & N/A & NDE$^b$ & U\\
\hline\\
Number & 48 & 20$^a$ & 13 & 5 & 4 & 23 &85 & 30\\
\hline

\hline	

\end{tabular}	
\label{Tab:numbers}	
\tablefoot{{\it a}) 12 of them also fall in the RH category. {\it b}) for 53 of them we have upper limits (see Paper II)}
\end{center}
\end{scriptsize}
\end{table}

\section{Fitting procedure}
\label{sec:correlation}

\begin{figure*}
\centering
\includegraphics[scale=0.5]{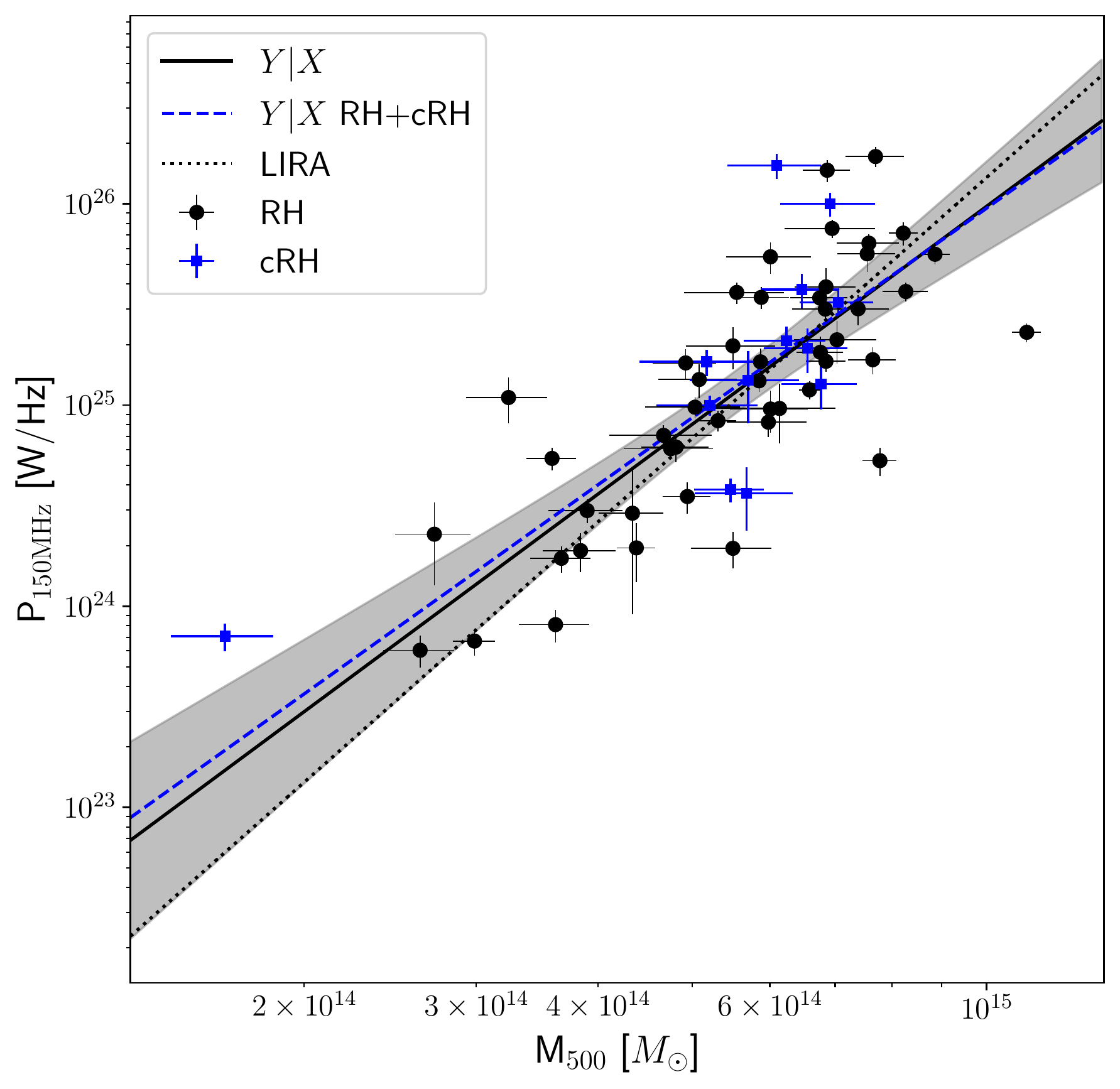}
\includegraphics[scale=0.5]{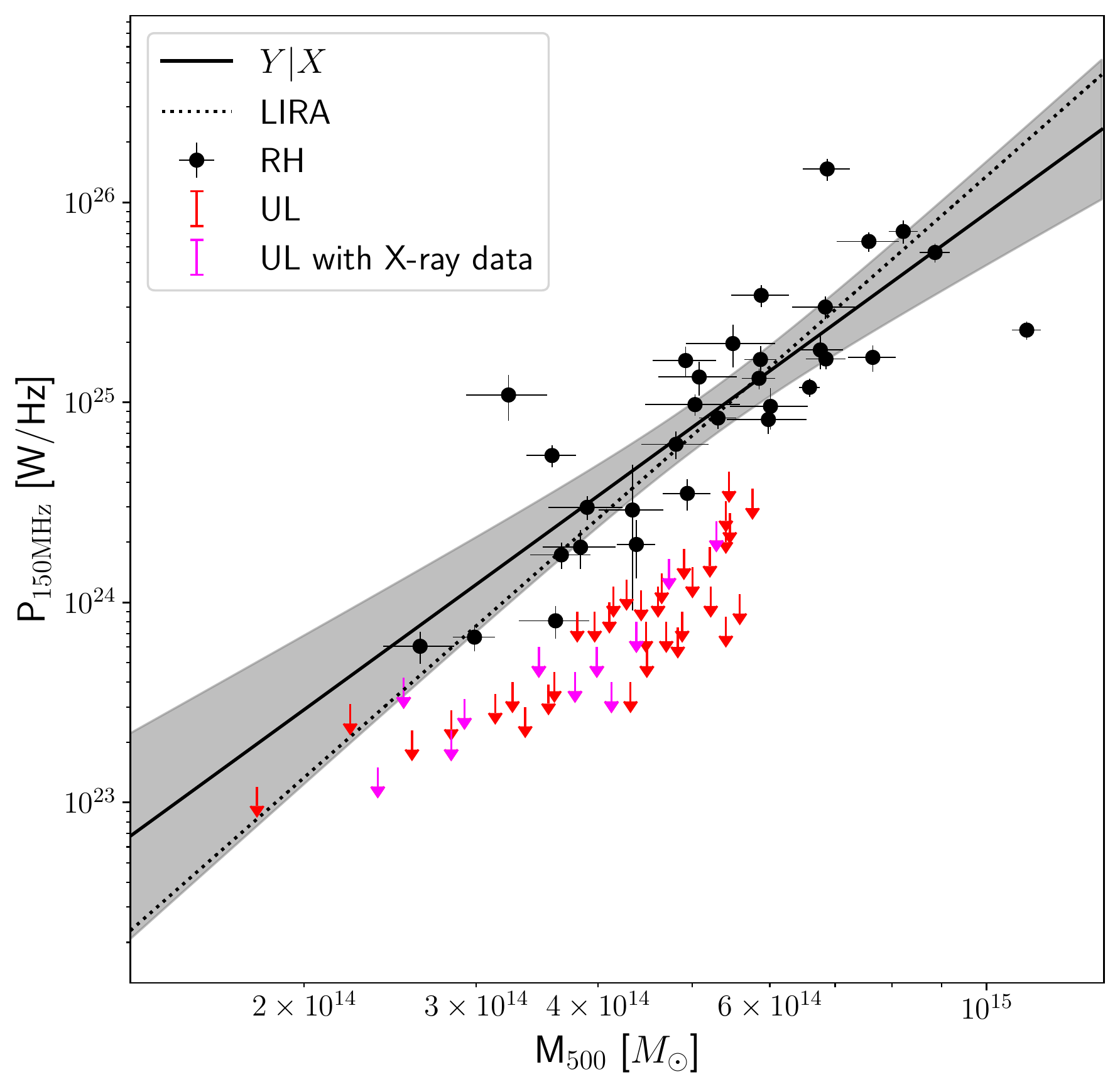}
\caption{Mass--radio power diagram. Left: Mass--radio power diagram for radio halos (black points) and candidate radio halos (blue squares) above the 50\% {\it Planck} completeness. The solid line and the shadowed region represent the correlation to radio halos only obtained with the BCES Y|X method and the 95\% confidence region. The dashed line is the correlation obtained considering also candidate radio halos. The black dotted line is obtained fitting only radio halos with LIRA bisector method. Right: Mass--radio power diagram for clusters with radio halos and upper limits with $0.06<z<0.4$ and $100\:\rm{kpc}<r_e<400\:\rm{kpc}$. The solid line and the shadowed region represent the correlation obtained with the BCES Y|X method and the 95\% confidence region, while the dotted black line is obtained with LIRA.}
\label{fig:P_M_compl50}
\end{figure*}

The resulting distribution of radio halos in the radio mass--power diagram is shown in Fig.~\ref{fig:P_M_compl50} (left). The uncertainties take into account the statistical, systematics and subtraction error. We refer the reader to Paper I for the estimation of these uncertainties. To derive the parameters of the correlation, we follow the fitting procedure outlined in \citet{cassano13} and \citet{cuciti21b}. Specifically, we adopt the Bivariate Correlated Errors and intrinsic Scatter (BCES) linear regression algorithms \citep{akritas96} to fit the observed $M_{500}-P_{150 \rm{MHz}}$ data points with a power-law in the form:

\begin{equation}
\mathrm{log}\left(\frac{P_{150 \rm{MHz}}}{10^{24.5}\mathrm{W/Hz}}\right)=B~\mathrm{log}\left(\frac{M_{500}}{10^{14.9}\,M_\odot}\right)+A , \label{eq:Pot-M1}
\end{equation}
where $A$ and $B$ are the intercept and the slope of the correlation, respectively.

Considering $Y=\log(P_{150 \rm{MHz}})-24.5$ and $X=\log(M_{500})-14.9$, and having $N$ data points ($X_i,Y_i)$ with errors $(\sigma_{X_i},\sigma_{Y_i})$, we estimate the raw scatter as:

\begin{equation}
\label{eq:scatter}
\sigma_{raw}^2=\dfrac{1}{N-2}\sum_{i=0}^N w_i (Y_i-BX_i-A)^2 ,
\end{equation}
where
\begin{equation}
\label{eq:weights}
w_i=\dfrac{1/\sigma_i^2}{(1/N)\sum_{i=0}^N1/\sigma_i^2}~~~~\mathrm{and}~~~~\sigma_i^2=\sigma_{Y_i}^2+B^2\sigma_{X_i}^2 .
\end{equation}

To evaluate the 95\% confidence region of the correlation, i.e. the area that has a 95\% probability of containing the `true' regression line, we calculated the 95\% confidence interval of the mean value of $Y$, $\left\langle Y\right\rangle $. For a given $X$, this is $\left\langle Y\right\rangle\pm\Delta Y $, where:

\begin{equation}
\Delta Y=\pm1.96 \sqrt{\bigg[\sum_{i=0}^N\dfrac{(Y_i-Y_m)^2}{N-2}\bigg]\bigg[\dfrac{1}{N}+\dfrac{(X-X_m)^2}{\sum_{i=0}^N(X_i-X_m)^2}\bigg]} ,
\end{equation}
where $Y_m=BX_i+A$ and $X_m=\sum_{i=0}^NX_i/N$ for each observed $X_i$.

The parameters of the correlation were derived considering only clusters with radio halos. They are shown in the top panel of Table \ref{Tab:fit}, together with the results from a 5000 bootstrap
resampling analysis. The parameters that we inferred considering also candidate radio halos are listed in the second panel of Table \ref{Tab:fit}. We also report the raw scatter of the correlation, $\sigma_{raw}$, and the Spearman correlation coefficient, $r_s$, which is a measure of the monotonicity of the relationship between two variables. In both cases, we obtained a value of $r_s\sim 0.8$ which means that the correlation is monotonically increasing and the probability that such a correlation is due to chance (p-value) is $<10^{-11}$.


In addition, we derived the parameters of the correlation considering only radio halos with the Bayesian LInear Regression in Astronomy method \citep[LIRA,][]{sereno16}. We used a mass distribution that depends on the redshift. The mass distribution of the {\it Planck} clusters we considered is limited, at large masses, by the steepness of the halo mass function and, at small masses,by the adopted 50\% mass completeness threshold. These two factors make the distribution approximately lognormal \citep{lima05}. We could then model the mass distribution of the observed sample as a lognormal function with redshift dependent mean and dispersion, which was shown to be a reliable approximation for the {\it Planck} sample \citep[see Fig. 12 in ][]{sereno15}
Moreover, we considered that both the radio power and the mass have a scatter with respect to a ``true'' mass, which is unknown. With this approach we are taking into account that the {\it Planck} masses, derived from the SZ signal, are proxies of the ``true'' mass. In general, we find a good agreement between the results of LIRA and BCES. With the assumption that Y and X are both linear proxies of a third variable, the fit is symmetric with respect to Y and X, in fact the LIRA results are very similar to the BCES bisector method. 

\begin{table} 
\begin{footnotesize}
\begin{center}
\caption{Fitting parameters}
\begin{tabular}{lcccccc}
\hline
\hline		
Method	&	B	&	err B	&	A	&	err A &$\sigma_{raw}$ & r$_s$\\
\hline\\
\multicolumn{7}{c}{Radio halos only}\\
\hline
BCES Y|X    	& 3.59	& 0.48	& 1.1	& 0.09& 0.39 & 0.79\\
Bootstrap		& 3.63	& 0.50	& 1.1	& 0.09& &\\
BCES bisector	& 4.30	& 0.10	& 1.2	& 0.06 & &\\
Bootstrap		& 4.35	& 0.50	& 1.2	& 0.09& & \\
BCES orthogonal	& 5.24	& 0.50	& 1.4	& 0.1 &  &\\
Bootstrap		& 5.37	& 0.50	& 1.4	& 0.1& & \\
LIRA    	& 4.30 	& 0.68	& 1.2	& 0.1 &  & \\

\\
\multicolumn{7}{c}{Radio halos and candidate radio halos}\\
\hline
BCES Y|X    	& 3.45	& 0.44	& 1.1	& 0.09& 0.39 & 0.78\\
Bootstrap		& 3.55	& 0.49	& 1.1	& 0.09& &\\
BCES bisector	& 4.19	& 0.16	& 1.3	& 0.05 &  &\\
Bootstrap		& 4.28	& 0.50	& 1.3	& 0.09& & \\
BCES orthogonal	& 5.19	& 0.69	& 1.4	& 0.1 &  &\\
Bootstrap		& 5.32	& 0.71	& 1.4	& 0.1& & \\
\\
\multicolumn{7}{c}{$0.06<z<0.4$ and $100 \mathrm{kpc}<r_e<400\mathrm{kpc}$}\\
\hline
BCES Y|X    	& 3.55	& 0.60	& 1.1	& 0.1& 0.35 & 0.86\\
Bootstrap		& 3.59	& 0.60	& 1.1	& 0.1& &\\
BCES bisector	& 4.11	& 0.10	& 1.2	& 0.07 &  &\\
Bootstrap		& 4.12	& 0.50	& 1.2	& 0.1& & \\
BCES orthogonal	& 4.79	& 0.50	& 1.3	& 0.1 &  &\\
Bootstrap		& 4.81	& 0.60	& 1.3	& 0.1& & \\
LIRA            & 4.13  & 0.70  & 1.2  & 0.14 & \\
\hline	
\hline

\end{tabular}	
\label{Tab:fit}	
\end{center}
\end{footnotesize}
\end{table}

\section{Scatter of the correlation}
\label{Sec:scatter}

The correlation between the radio power of radio halos at 150 MHz and the mass of the host clusters shows a relatively large scatter (Fig. \ref{fig:P_M_compl50}). While measurement errors play a role, there is likely some inherent variation due to the complex mixture of clusters in different dynamical stages and radio halos with different spectra and different sizes \citep{cuciti21b}. To ascertain the role of dynamical status on the scatter of the correlation, similar to \citet{cuciti21b}, we used the X-ray morphological disturbance as a proxy. This is a single parameter, derived from the combination of the concentration \citep[$c$,][]{santos08} and centroid shift \citep[$w$,][]{bohringer10} parameters.

\begin{figure*}
\centering
\includegraphics[scale=0.4,trim={0cm 0cm 0cm 0cm},clip]{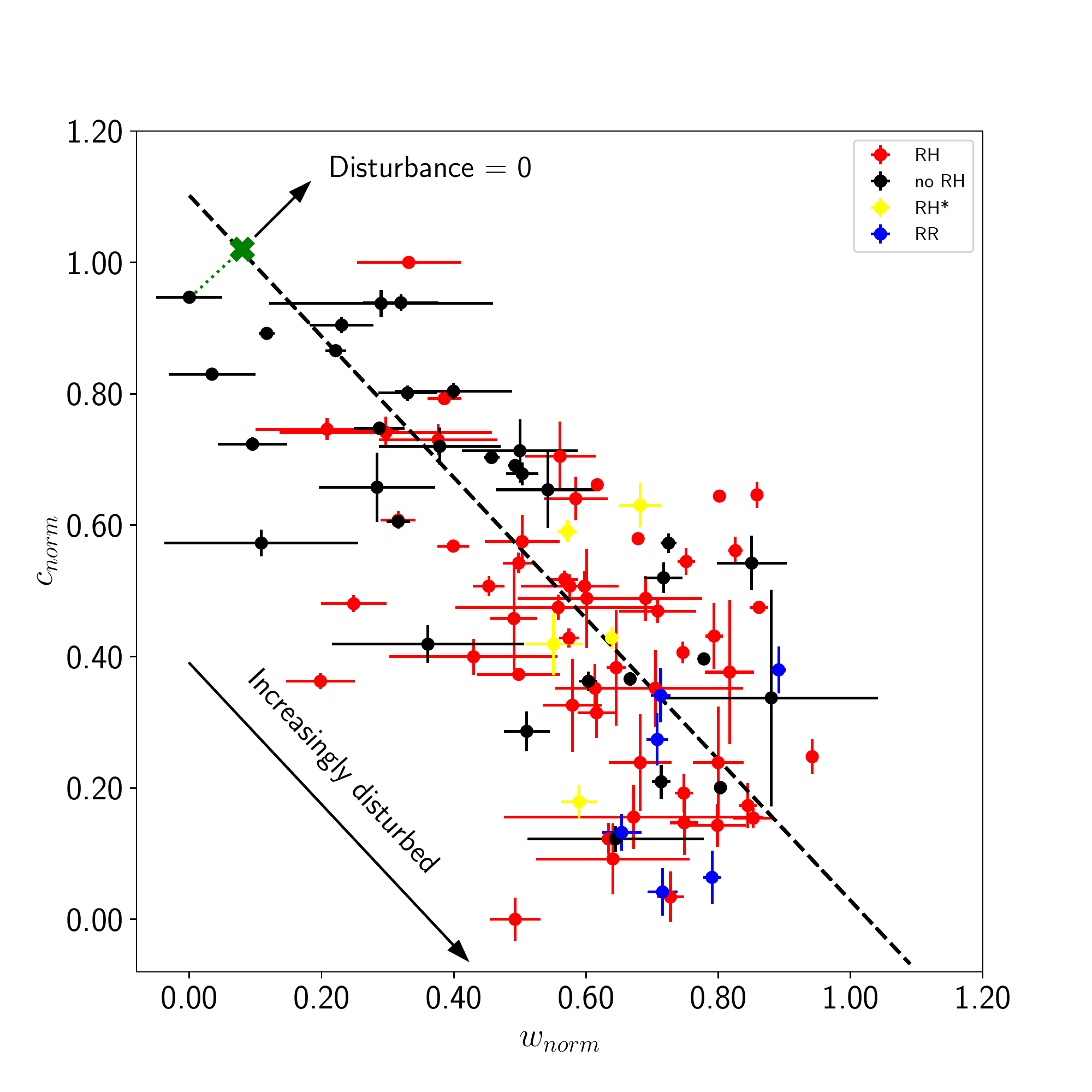}
\includegraphics[scale=0.4,trim={0cm 0cm 0cm 0cm},clip]{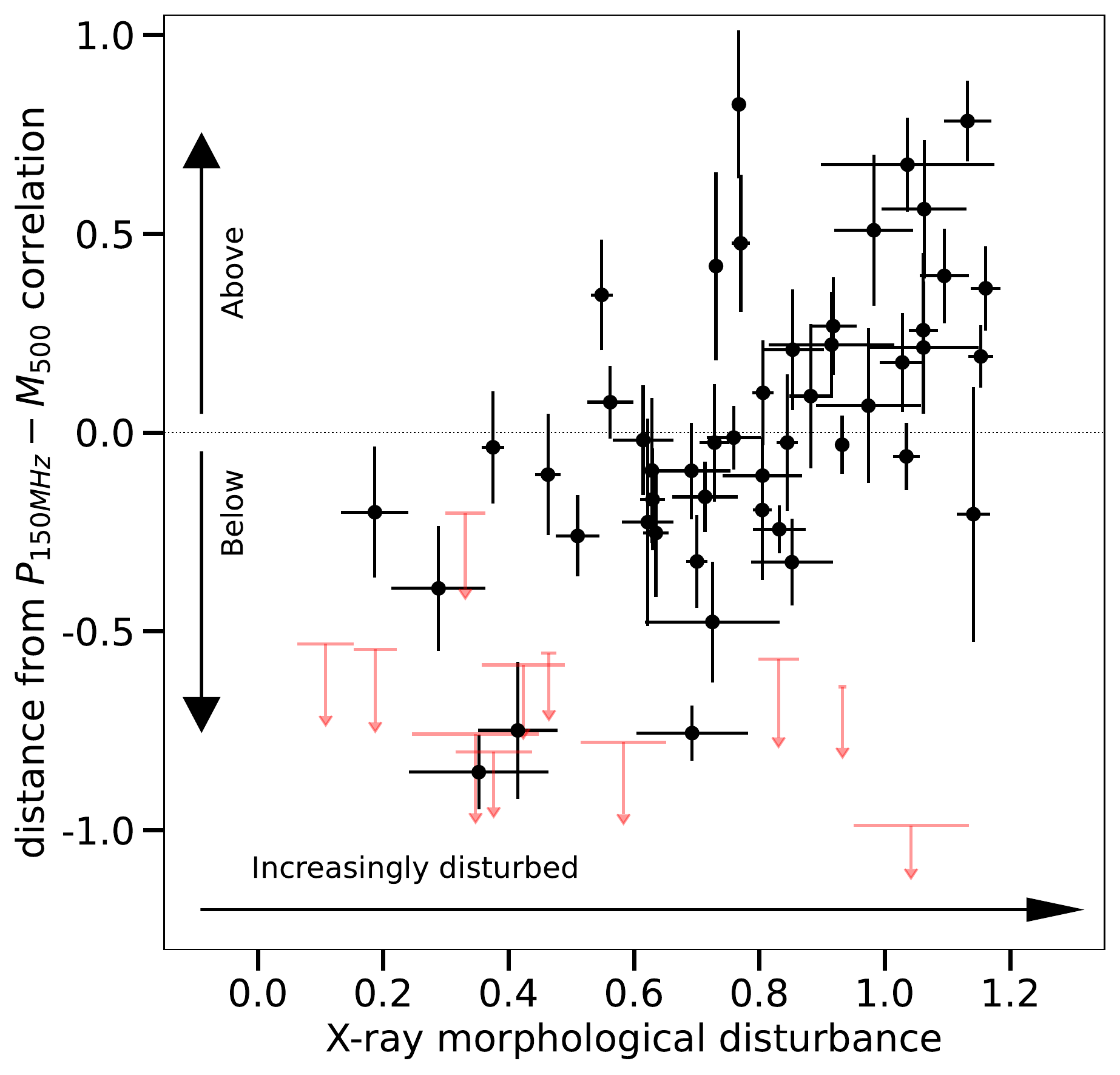}
\caption{X-ray morphological disturbance. Left: Normalised concentration parameter--centroid shift diagram for all the clusters of the sample with available X-ray observations. The black dashed line is the regression line. The green cross marks the projection on the regression line of the cluster with disturbance = 0. Right: Distance of radio halos (black points) and upper limits (red arrows) from the mass--radio power correlation (BCES Y|X method) vs. X-ray morphological disturbance.}
\label{fig:scatter_PM}
\end{figure*}

Among the 309 clusters of the sample, 143 (46\%) have X-ray (Chandra or XMM-Newton) data. We used these archival data to derive the morphological parameters $c$ and $w$ (Paper III).
Here, we adopt a slightly different approach to measure the X-ray morphological disturbance with respect to \citet{cuciti21b}. We start by normalising the values of each morphological parameter, $\mathcal{P}$, such that:

\begin{equation}
    \mathcal{P}_{\rm norm} = \frac{\mathrm{log}(\mathcal{P}_i) - \mathrm{min}(\mathrm{log}(\mathcal{P}))}{\mathrm{max}(\mathrm{log}(\mathcal{P})) -\mathrm{min}(\mathrm{log}(\mathcal{P}))} .
\end{equation}
This allows us to take into account the different ranges of values covered by $c$ and $w$. We then fit the distribution of $c_{\rm norm}$ versus $w_{\rm norm}$ (Fig.~\ref{fig:scatter_PM}, left) with a power law in the form:

\begin{equation}
  c_{\rm norm} = m\times w_{\rm norm} +q  ,
\end{equation}
using the same methods described in Section \ref{sec:correlation}. With the BCES bisector method we obtained $m=-1.07\pm0.07$ and $q=1.1\pm0.04$.
We derived the projected position of each cluster along the fitting line and we assumed that the cluster with X-ray disturbance=0 is the first along the line starting from the top left corner of the plot (i.e. high $c$ and low $w$, see Fig. \ref{fig:scatter_PM}, left). The disturbance of the other clusters is calculated as the distance along the fitting line from the cluster with disturbance=0.
The choice of the zero point is arbitrary, and hence the values of the disturbance do not have a direct physical meaning if not in comparison with the other clusters. 
With this definition higher values of the X-ray morphological disturbance indicate clusters that are more dynamically active. We note that with this definition, a cluster with a pronounced core (high $c$) that is very disturbed on the large scale (high $w$), would have similar disturbance to a cluster which appears less morphologically disturbed (lower $w$) but has a disrupted core (lower $c$).

We show in Fig.~\ref{fig:scatter_PM} the distance (on the $P_{150 \rm{MHz}}$ axis) of radio halos from the mass--radio power correlation vs their X-ray morphological disturbance.
We confirm the trend between these two quantities, with radio halos that lie above the correlation being located in more dynamically disturbed clusters. Fig.~\ref{fig:scatter_PM} suggests that the merger activity has a key role in determining the position of radio halos with respect to the correlation, thus inducing at least part of the large scatter. This can be explained by two aspects, or most likely, by the combination of them: the different time scales of the mergers and the different types of merger. This is supported by numerical simulations showing that the emission of radio halos increases in the early stages of the
merger when turbulence accelerates electrons, and then decreases along with the dissipation of turbulence at later merger stages \citep{donnert13}. In particular, in one simulated cluster with $M_{200}=10^{15}M_\odot$, \citet{donnert13} showed that in the initial stage of the merger the radio power increases of a factor of $\sim30$ on a timescale of $\sim1.5$ Gyr and then decreases of the same amount in less than 1 Gyr. In the mass--radio power plane this translates into
a migration of clusters from the region of the upper limits to the correlation (or above) and a progressive movement back. On the other hand, the most powerful mergers (high mass ratio and/or small impact parameter) dissipate a larger amount of energy in the ICM and therefore the free energy available for particle acceleration and magnetic field amplification is higher \citep[eg][]{cassano16}. 

In general, it is expected that the clusters for which we derived upper limits on the radio power are less disturbed. Unfortunately, while we have X-ray observations for all the radio halos in the sample analysed in this paper, we only have X-ray data for 11 clusters with radio upper limits. We plot these upper limits in Fig.~\ref{fig:scatter_PM} (right) as red arrows. Although the incompleteness of the X-ray information does not allow us to derive meaningful statistical conclusions (see Paper IV for a deeper discussion on this), it is interesting to note that $\sim70\%$ of the upper limits have low X-ray morphological disturbance. We did not find evidence of a relation between the disturbance of the clusters with upper limits and their mass and/or redshift, however we stress again that larger numbers are necessary to make this kind of consideration.  

In the literature there are other parameters that combine two morphological estimators, such as $c$ and $w$ \citep[e.g.][]{rasia13,lovisari17, campitiello22, ghirardini22}. One of these is the parameter $M$, first introduced by \citet{rasia13} and then used in other works \citep[e.g.][]{lovisari17, campitiello22}. 
To double check our results, in Appendix \ref{app:rasia} we derive the $M$ parameter for the clusters of the sample and compare it with the value of the disturbance. We found that the two values are very well correlated. This indicates that there is agreement in the classification of the dynamical state of clusters using these two methods. In fact, we show that even considering $M$ to classify the dynamics of clusters, we find evidence that the most disturbed clusters are scattered up with respect to the correlation.

\section{On the distribution of clusters in the $M_{500}-P_{\rm 150\,MHz}$ diagram}
\label{sec:bimodality}

It is known that massive galaxy clusters ($M_{500}>5-6\times 10^{14}M_\odot$) have a bimodal behaviour in the $M_{500}-P_{\rm 1.4\,GHz}$ diagram, with disturbed clusters populating the correlation and relaxed systems appearing as a separated population with upper limits to their radio power \citep{cassano13, cuciti21b}. Here we investigate whether a similar behaviour is found also at 150 MHz. To this end, we focus on clusters at $0.06<z<0.4$ and on radio halos with 100 kpc$<r_e<$400 kpc (Fig.~\ref{fig:P_M_compl50}, right). In this ranges we have 32 radio halos and 46 upper limits. In this redshift range current observations are able to put stringent upper limits (Paper II). The range of $r_e$ is chosen because upper limits are all done with $r_e=200$ kpc (which is consistent with the mean $r_e$ of radio halos in the sample, see Paper II), therefore the radio power of radio halos that are more than a factor of two larger or smaller than this value would not be easily comparable with these upper limits. Fig. \ref{fig:P_M_compl50} shows that we can now reliably constrain the correlation down to masses of the order of $\sim 3\times 10^{14}M_\odot$, whereas below we do not have sufficient statistics. 
To study the distribution of clusters with and without radio halos in the $M_{500}-P_{\rm 150\, MHz}$ diagram, we adopt two approaches, that we outline below.

\subsection{Akaike Information Criterion}
In this Section we use the Akaike information criterion \citep[AIC][]{akaike74} to determine if the distribution of clusters in the $M_{500}-P_{\rm 150\, MHz}$ plane is better described with a model composed of two separate populations, one for the radio halos and one for the upper limits, or if it is better described by a single population. The AIC value of each model is calculated as:
\begin{equation}
    \mathrm{AIC} = 2k - 2 \mathrm{ln}(\hat{L}) ,
    \label{eq_aic_1}
\end{equation}
where $k$ is the number of free parameters of the model and $\hat{L}$ is the maximum value of the likelihood function as a function of the model parameters, see Eq.~(\ref{eq_aic_2}). In general, given a set of models, the preferred model is the one with the minimum AIC value. Differently from the full LIRA regression analysis adopted in Section \ref{sec:correlation}, here we did not consider the scatter of the {\it Planck} mass with respect to the true mass, and the modelling of the mass distribution.For each scaling relation, there are three fitting parameters, i.e. the normalisation $A$, the slope $B$, and the scatter $\sigma$. We assumed that the value for a measurement for which only an upper limit is provided follows a uniform distribution spanning the range from $P_{UL}-1$ to $P_{UL}$, where $P_{UL}$ is the logarithm of the value of the upper limit reported in Paper II. After marginalisation over the not observed true values $Y$, the likelihood can be written as
\begin{eqnarray}
L & = & \prod_i^{N_\text{det}} 
\frac{1}{\sqrt{2 \pi (\sigma_\text{det}^2 + \delta_i^2)}} 
\exp
\left( 
\frac{(A_\text{det} + B_\text{det} x_i - y_i)^2}{\sigma_\text{det}^2 + \delta_i^2} 
\right) \nonumber \\
& \times & 
\prod_i^{N_\text{ul}} \frac{ 
\text{erf}\left(\frac{A_\text{ul} + B_\text{ul} x_i - Y_\text{min}}{\sqrt{2}\sigma_\text{ul}} \right)  - \text{erf}\left(\frac{A_\text{ul} + B_\text{ul} X_i - Y_\text{max}}{\sqrt{2}\sigma_\text{ul}} \right) 
}{2(Y_\text{max} - Y_\text{min})} ,
\label{eq_like_ul}
\label{eq_aic_2}
\end{eqnarray}
where $x_i$ and $y_i$ are the measured values of the $i$-th cluster, $\delta_i$ is the measurement uncertainty on $y_i$, $\text{erf}$ is the error function, $Y_\text{max} = P_{UL}$, and $Y_\text{min} = P_{UL} -1$ (see Appendix \ref{app:UL}). Uncertainties on the $x$ value are not considered in Eq.~(\ref{eq_like_ul}).

In the first model we fitted both radio halos and upper limits with a single scaling relation, i.e. $A_\text{det} = A_\text{ul}$, $B_\text{det} = B_\text{ul}$, and $\sigma_\text{det} = \sigma_\text{ul}$. In this case $k=3$. In the second model, we fitted radio halos and upper limits with two separate scaling relations. In this case $k=6$.

We obtained AIC = 150.9 for the first model and AIC = 46.2 for the second model.
According to the AIC, the model with two separate correlations, one for the radio halos and one for the upper limits is the one that minimises the information loss and better describes the distribution of clusters in the mass--radio power diagram.
On the other hand, the model with just one correlation is $\sim 2 \times 10^{-23}$ times less probable than the model with two relations to minimise the information loss. 

To check for agreement between the maximum likelihood analysis based on Eq.~(\ref{eq_like_ul}) and the Bayesian analysis based on LIRA in the case of RHs and ULs following two different relations, we resampled the data 1000 times and fitted each mock data-set with LIRA under the same assumptions detailed in Section \ref{sec:correlation}. Detected radio halos were extracted from Gaussian distributions centred around the measured value of the radio power and with dispersion equal to the associated uncertainty. The value of each upper limit was extracted from a uniform distribution spanning the range from $P_{UL}-1$ to $P_{UL}$. Each sampled population was fitted with LIRA under the same regression scheme detailed in Sec.~\ref{sec:correlation}. Posterior distributions were obtained putting together the mean estimated values of the marginalised distributions of each fitted sample. The best fit values from the maximum likelihood analysis are in full agreement with the Bayesian analysis exploiting LIRA (Section \ref{sec:correlation}).

\subsection{Monte-Carlo test}
\label{sec:MC}
\begin{figure}
\centering
\includegraphics[width=\hsize]{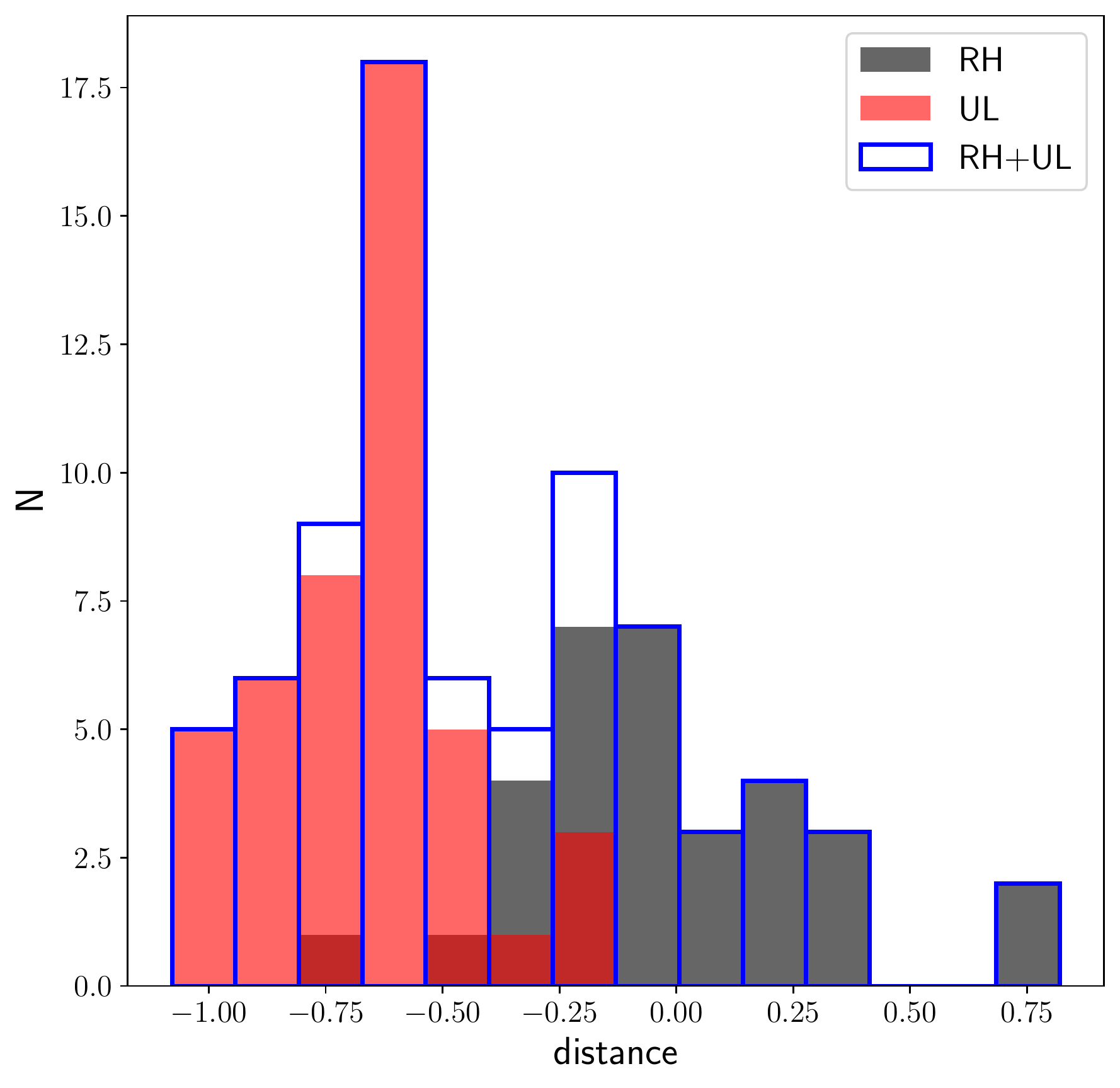}

\caption{Distribution of distances of radio halos (black) and upper limits (red) from the correlation shown in Fig.~\ref{fig:P_M_compl50} (right). The sum of upper limits and radio halos in each bin is given by the blue histogram.}
\label{fig:histo}
\end{figure}

In this Section we focus on the distance of clusters from the correlation. In Fig.~\ref{fig:histo} we show the distribution of the distance along the $P_{\rm 150\,MHz}$ axis of radio halos and upper limits from the correlation shown in Fig.~\ref{fig:P_M_compl50}, right. We note that radio halos and upper limits exhibit a bimodal distribution, albeit with some overlapping distances. Here we want to test whether the distances shown in Fig. \ref{fig:histo} are compatible with clusters being normally distributed above the correlation and uniformly distributed below the correlation. In fact, we cannot exclude that the peak of the distribution of the upper limits is driven by the sensitivity limit of our observations. The distribution of the radio power of clusters that currently have an upper limit may be broader and may extend to much larger distances from the correlation. The peak of the distribution of the radio halos is instead attributed to the fact that clusters with radio halos follow the correlation and are distributed around it with a certain scatter. In this view, the correlation represents a region of the diagram that is more densely populated than the rest. Here we use Monte-Carlo simulations to exclude the possibility that the region below the correlation is instead uniformly populated and that some of the less powerful radio halos are not detected because they are close to the sensitivity limit of our observations. In this case the decrease of number of radio halos below the correlation (the left hand side of the black distribution in Fig.~\ref{fig:histo}) would be also artificial. We point out that this scenario does not hold for clusters with $M_{500}>5.5 \times 10^{14}M_\odot$, because at those masses we have only well detected radio halos and no upper limits\footnote{The clusters for which we could not derive a meaningful upper limit are uniformly distributed in mass (Paper II), so this cannot be due to the missing upper limits}. This does not imply that the 100\% of the clusters at high mass have a radio halo because here we are not taking into account clusters with RR only, U cases and NDE clusters without an available upper limit. We refer the reader to Paper IV for the discussion on the fraction of clusters with radio halos as a function of mass. 

\begin{figure*}
\centering
\includegraphics[scale=0.4]{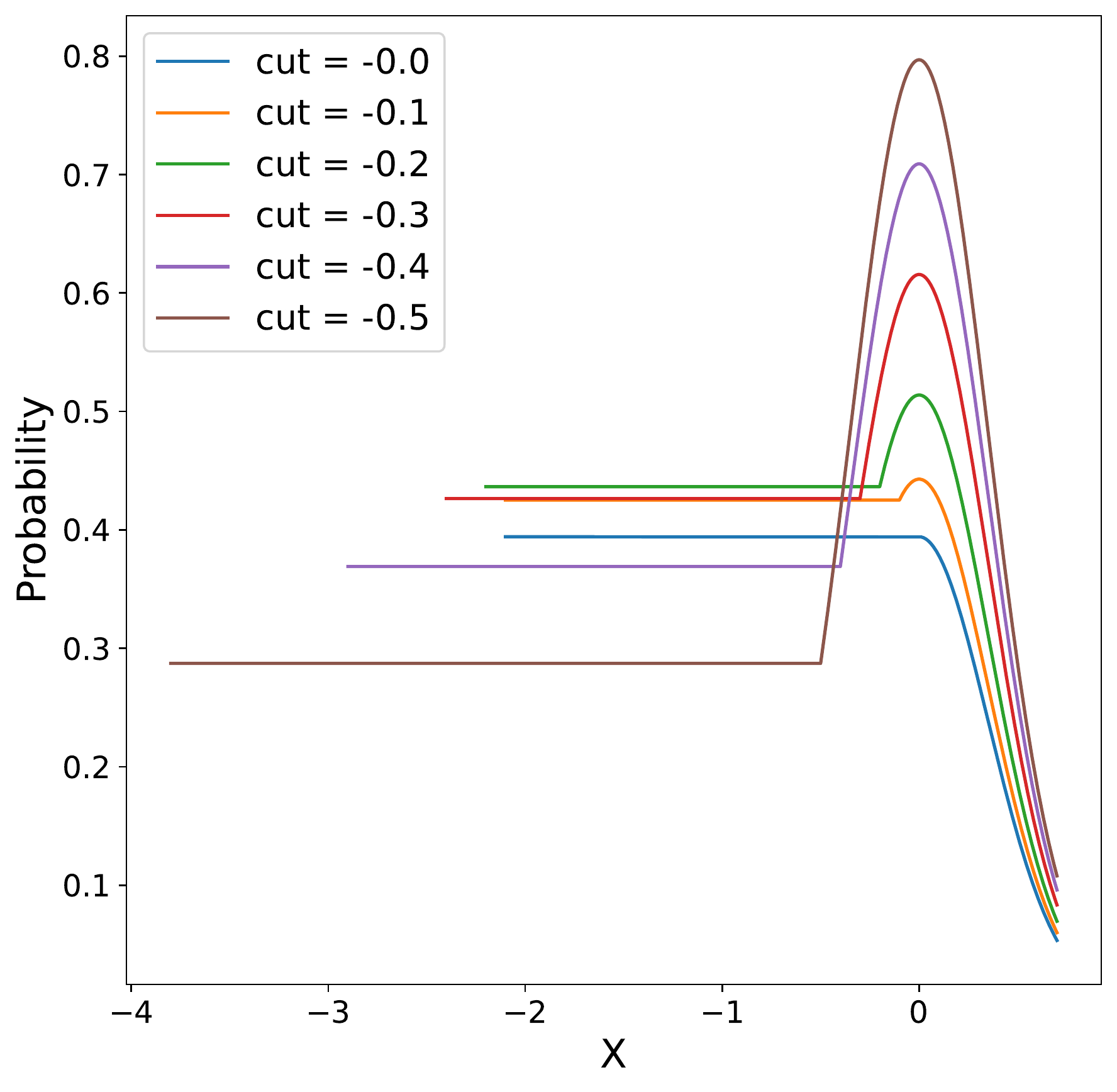}
\hspace{1cm}
\includegraphics[scale=0.4]{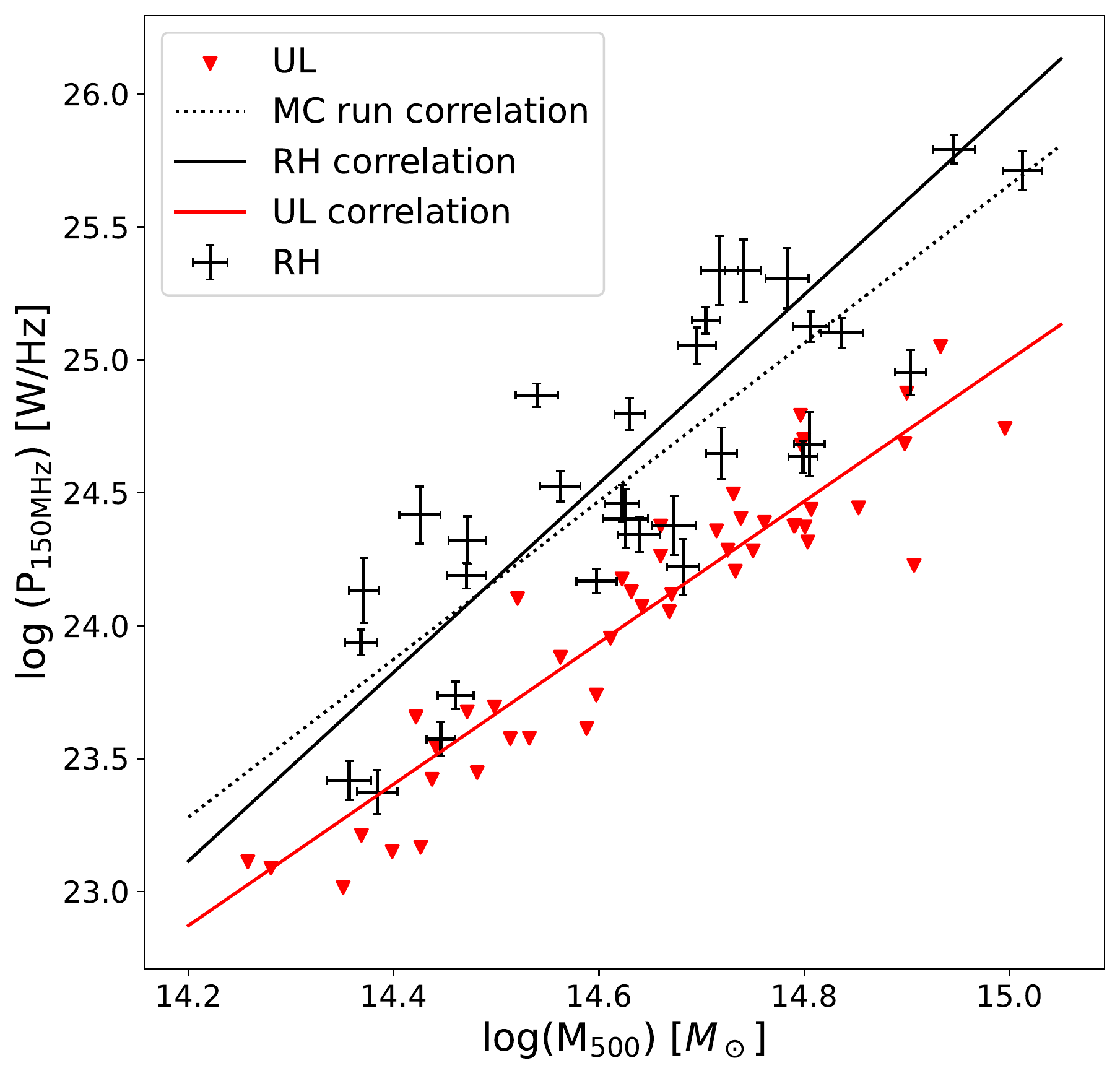}
\caption{Monte-Carlo test. Left: Probability density functions (PDFs) with different cuts. PDFs are normalised so that the area below each curve corresponds to 1. We used these PDFs to randomly extract values to add to the radio power of halos on the correlation and distribute them on the M-P plane. Right: an example of the random distribution of clusters in the mass--radio power diagram. Errorbars are radio halos, arrows are upper limits. The black and red solid lines represent the radio halo and upper limits correlation we found in the sample, respectively.}
\label{fig:MC}
\end{figure*}

Upper limits seem to have a trend in radio power vs mass. This trend is mainly driven by the fact that the clusters at the highest redshifts are also among the most massive ones and, even if the upper limits are similar in terms of flux density to the lower redshift ones, their radio power is higher. As a first step, we evaluate this correlation in the form $\mathrm{log}(P_{UL}) = m\times \mathrm{log}(M_{UL}) +q$. We obtained $m=2.66$ and $q=-14.9$, with a scatter $\sigma_{raw}=0.2$ dex estimated as in eq. \ref{eq:scatter}. In our sample we have 78 clusters with $0.06<z<0.4$ that have either a radio halo with $100\,\mathrm{kpc}<r_e<400$ kpc (32) or an available upper limit (46). We started generating 78 clusters with random masses, with the requirement that the distribution of masses resembles the one we have in the sample. To do so, we divided the mass range into three bins and for each bin we distributed the same number of clusters that we have in the sample also in the simulated sample. We assign a radio power to these clusters so that they are distributed according to the radio halos correlation in the $M_{500}-P_{\rm 150\,MHz}$ diagram. 
Then we added a random value to each radio power, extracted from one of the probability density functions (PDFs) shown in Fig. \ref{fig:MC} (left). The zero point on the X-axis of Fig. \ref{fig:MC} (left) represents the correlation. If a negative value of X (on the left of the zero point) gets randomly extracted, the cluster gets shifted below the correlation and vice versa. Around the correlation, clusters are normally distributed with $\sigma = \sigma_{raw} =0.35$ dex. 
Below the correlation clusters are uniformly distributed. The different PDFs shown in Fig. \ref{fig:MC} (left) correspond to different ``cuts'', i.e. different distances from the correlation where the uniform distribution starts. In this formalism, cut$=-0.0$ means that clusters are uniformly distributed from the correlation down, cut$=-0.1$ means that clusters are normally distributed until 0.1 dex below the correlation and then they are uniformly distributed further below.   

Given this initial random distribution on the $M_{500}-P_{\rm 150\,MHz}$ diagram, we decide whether a cluster is classified as radio halo or upper limit based on its position: if it is above the upper limits correlation, then it is a radio halo, otherwise it is an upper limit. We define how far below the correlation clusters can be distributed by imposing that the number of radio halos and upper limits in the simulation are similar to the ones we have in the sample (32 radio halos and 46 upper limits). This sets the lower limit to the left of the PDFs in Fig. \ref{fig:MC} (left). Since the upper limits distribution represents our observational limit, when a cluster is classified as an upper limit, we assign a random value to its radio power within the scatter of the upper limits correlation. This is to take into account the fact that there is some scatter in the values of upper limits related to the quality of the images, the presence of bright sources in the field and the cluster redshift, therefore some upper limits lie above and some below the upper limits correlation.

We repeated this procedure 10000 times and each time we computed the radio halo correlation. The mass--radio power diagram resulting from one of the runs with cut = 0.0 is shown in Fig.~\ref{fig:MC} (right) as an example. 
We evaluated the probability that, over the 10000 Monte-Carlo runs, the fraction of radio halos falling within the scatter of the upper limits correlation is smaller than the one we measure in the sample (0.03) or, in other words, that radio halos and upper limits are more separated than in the sample. We list the obtained probabilities for each cut in Table \ref{Tab:MC}.   


\begin{table} 
\begin{small}
\begin{center}
\caption{Results of Monte-Carlo test}
\begin{tabular}{c|c}

cut &	probability	\\
\\
$-0.0$ & 0.30 \% \\
$-0.1$ & 0.33 \% \\
$-0.2$ & 0.57 \% \\
$-0.3$ &  1.02 \% \\
$-0.4$ & 2.30 \%\\
$-0.5$ & 4.65 \% \\

\end{tabular}	
\label{Tab:MC}	

\end{center}
\end{small}
\end{table}

These results suggest that the probability that the observed distribution of radio halos and upper limits in the mass--radio power diagram is due to our sensitivity limit applied to a uniform distribution of clusters starting below an upper envelope (cut = $0.0$) is $<0.3$\%. This probability is still $<5\%$ when the uniform distribution starts 3 times below the correlation (cut=$-0.5$). This means that the correlation we observe in our sample is indeed a more densely populated region of the $M_{500}-P_{\rm 150\,MHz}$ diagram and that the decrease of radio halos below the correlation in Fig. \ref{fig:histo} is not simply due to the sensitivity limit of our observations. 
In addition to this, we stress again that the distribution of massive clusters in the test (e.g. Fig.~\ref{fig:MC}, right) is significantly different from what we see in the sample (Fig.~\ref{fig:P_M_compl50}, right), where all clusters with $M_{500}>5.5 \times 10^{14}M_\odot$ host radio halos.

\section{Comparison with high frequency samples}
\label{sec:comparison}
\begin{figure*}
\centering
\includegraphics[scale=0.5]{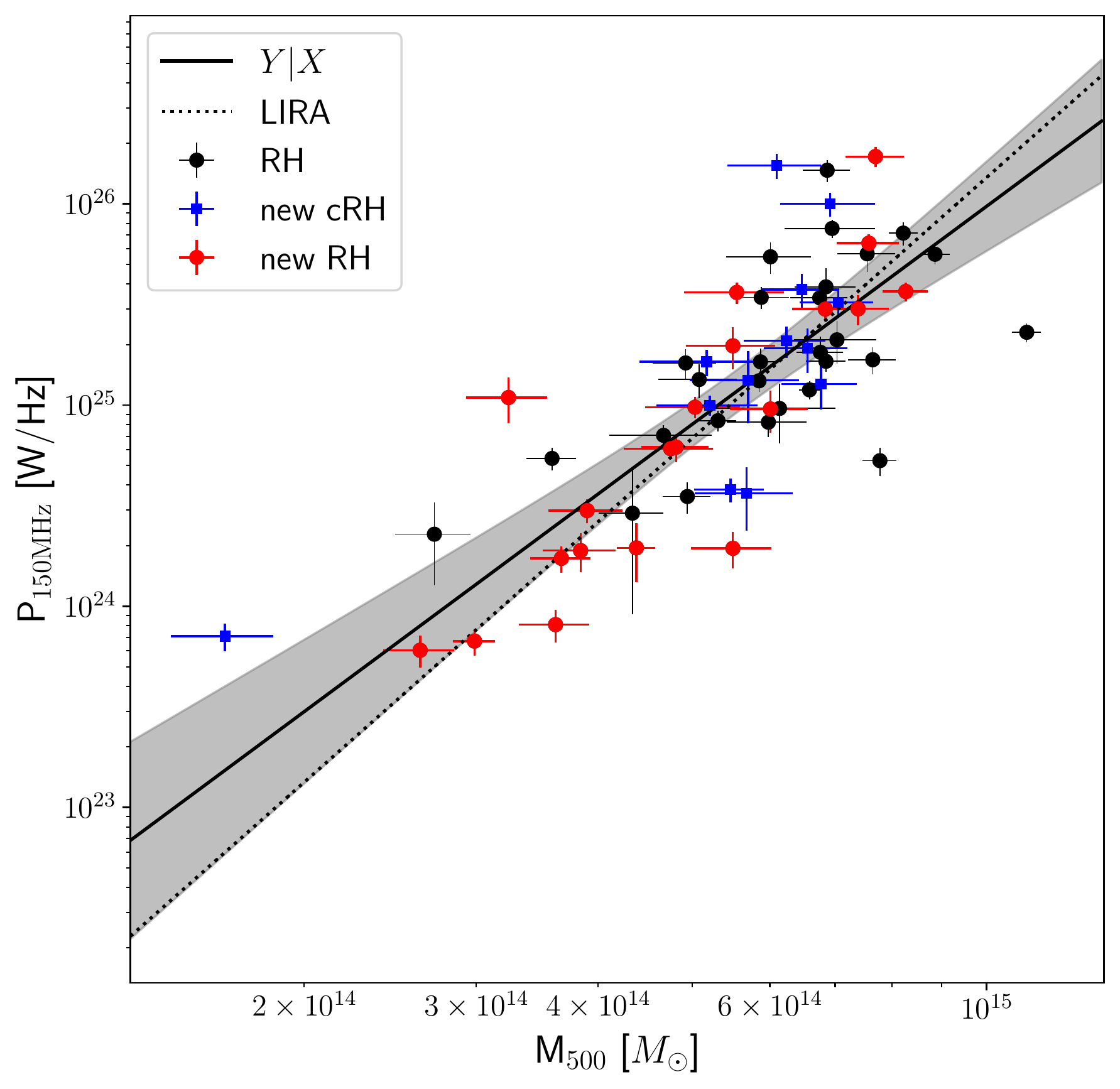}
\includegraphics[scale=0.5]{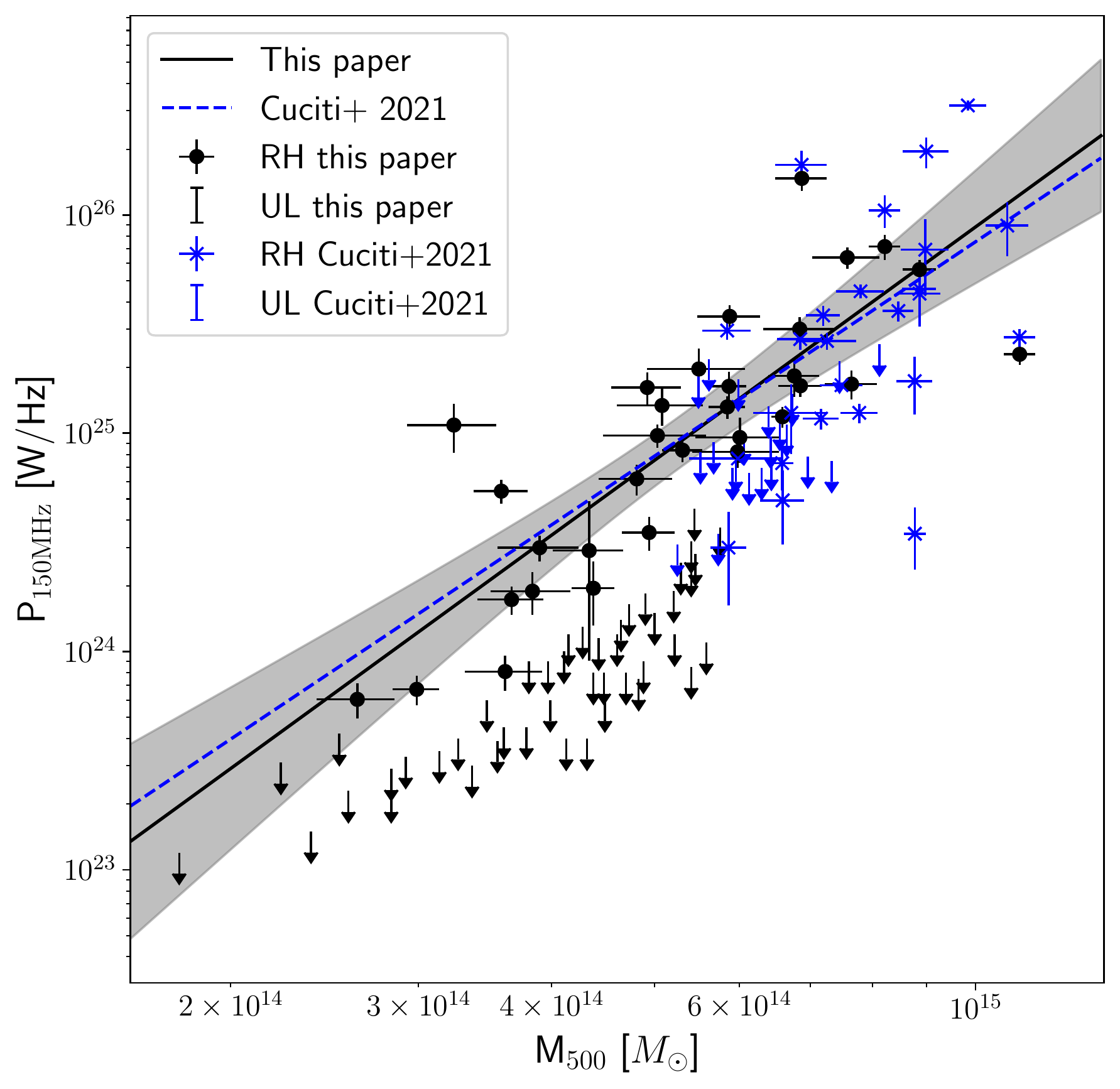}

\caption{Comparison with previous studies. Left: black points represent radio halo clusters that where already known in the literature, red points are newly discovered radio halos, blue squares are new candidate radio halos. Right: Comparison with \citet{cuciti21b} re-scaled at 150 MHz assuming $\alpha=-1.3$. Black points and arrows are radio halos and upper limits from the LOFAR sample, blue stars and arrows are radio halos and upper limits in \citet{cuciti21b}. The black solid line is the same as in Fig.~\ref{fig:P_M_compl50} (right). The dashed blue line is the correlation obtained for the \citet{cuciti21b} sample including radio halos and USSRH.}
\label{fig:GMRT}
\end{figure*}

In Fig.~\ref{fig:GMRT}, we show a comparison between the sample that we analysed here and previous studies. On the left panel, we highlight the radio halos of the LOFAR DR2 {\it Planck} sample that were already reported in the literature (black points). The number of radio halos is now almost two times larger than in previous studies and almost three times larger if we consider also candidate radio halos. Specifically, among the confirmed 48 radio halos analysed in this paper, 20 are new discoveries. The majority of the new detections are located at masses below $5\times 10^{14}M_\odot$. At these masses, turbulent re-acceleration models predict that a large fraction of clusters should host USSRHs (Paper IV). However, the available observations at higher frequencies do not allow us to test this expectation. In fact, for most of the low mass clusters of the sample, the expected flux density at 610 MHz with $\alpha=-1.3$ would be comparable to the typical upper limits found with the GMRT at those frequencies \citep{venturi08, kale15, cuciti21a}. In this respect, we have been recently allotted uGMRT observations (PI R. Cassano) of the radio halos discovered with LOFAR which will be 5--10 times more sensitive than those used in \citet{cuciti21a} and will allow us to detect radio halos with spectra as steep as $\alpha\sim -1.5-2$. 

In the right panel of Fig.~\ref{fig:GMRT} we show both radio halos and upper limits from the LOFAR DR2 {\it Planck} sample together with those from the sample analysed in \citet{cuciti21a}, which was up to now the largest complete sample of clusters with deep radio observations at frequencies larger than 600 MHz. We plot both radio halos and USSRH (and candidate USSRHs) from the GMRT sample. For consistency, we also plot the correlation derived for radio halos plus USSRHs for the statistical sample \citep[see Table 1 in][]{cuciti21a}. We note that the correlation in \citet{cuciti21b} was derived at 1.4 GHz. We assumed a spectral index $\alpha=-1.3$. The right panel of Fig.~\ref{fig:GMRT} shows that we can now investigate the correlation and the position of the upper limits with respect to the correlation in a much larger mass range, that was previously inaccessible. In spite of the low statistics and relatively small mass range of the high frequency sample, we note that the correlation found at 150 MHz is well in line with the extrapolation of the one derived at 1.4 GHz also in the low mass regime. From a theoretical point of view, the scatter of the correlation is expected to increase at low frequency because of the intervening population of USSRHs that are less luminous than classical radio halos \citep{cassano10b}. With respect to the correlation derived in \citet{cuciti21b} at 1.4 GHz, which has a scatter of $\sim0.3$ dex, at 150 MHz we find a larger scatter ($\sim 0.4$).  

Not only the number of radio halos, but also the number of available upper limits has dramatically increased. Most importantly, the upper limits derived with LOFAR are deeper than those derived at high frequency (see Fig. 13 in Paper II for a comparison). For clusters with similar masses ($M_{500}\sim 6\times 10^{14}M_\odot$) the upper limits at 150 MHz are typically a factor of 2--5 lower than those at 1.4 GHz (again assuming $\alpha=-1.3$). While at 1.4 GHz, some upper limits below $M_{500}\sim 6.5\times 10^{14}M_\odot$ were close or consistent with the correlation, at 150 MHz this takes place only for $M_{500}<3\times 10^{14}M_\odot$. This suggests the existence of a correlation extending at least down to masses of the order of $M_{500}=3\times 10^{14}M_\odot$.

\section{Summary and conclusions}

Statistical studies of diffuse radio emission in galaxy clusters have shown that radio halos are predominantly found in merging clusters \citep{cassano10, cuciti15, cuciti21b} and that the radio power of radio halos correlates with the mass of the host cluster \citep{basu12, cassano13, cuciti21b}. On the other hand, clusters without radio halos, typically associated with non-merging systems, populate a different region of the mass--radio power diagram, below the correlation. These findings support the turbulent re-acceleration model for the formation of radio halos \citep[e.g][]{brunetti01}, which predicts a heterogeneous population of radio halos with different spectral properties. While massive clusters in energetic mergers should have spectral index of $\alpha\sim -1.3$, a large fraction of less massive clusters or less energetic mergers should produce the so-called USSRHs ($\alpha< -1.5$), visible at low radio frequencies. The great majority of the statistical studies performed so far have used observations at $\sim$GHz frequencies, thus limiting the number of detected radio halos and the explorable mass range. A first analysis of the statistics of radio halos at 150 MHz has been presented in \citet{vanweeren21}, who analysed the {\it Planck} clusters in the first LoTSS data release. They collected a sample of 26 clusters, with 10 radio halos and 12 candidate radio halos. They investigated the mass--radio power correlation at 150 MHz, but their statistics was too low to allow for a solid comparison with higher frequency results.

In this paper, we analysed the PSZ2 clusters covered by the second data release of LoTSS \citep{botteon22} with the aim of characterising the distribution of clusters in the $M_{500}-P_{\rm 150\,MHz}$ diagram. We focused on the 221 clusters lying above the 50\% {\it Planck} mass completeness line (Fig.~\ref{fig:M_z_compl50}). These clusters span wide mass ($3-10\times10^{14}M_\odot$) and redshift ($0.02-0.6$) ranges. Among them we detected 49 radio halos (one is not considered here due to the unreliability of the flux density measurement) and 13 candidate radio halos. In addition, we derived upper limits to the radio power of 53 clusters (Paper II). 

We confirm the existence of a correlation between the radio power of radio halos and the mass of clusters also at 150 MHz (Fig.~\ref{fig:P_M_compl50}). We derived the parameters of the correlation with the BCES linear regression algorithms and with the Bayesian method LIRA. We applied these methods to radio halos only, radio halos plus candidate radio halos, and to a subsample of radio halos with $0.06<z<0.4$ and $100\,\mathrm{kpc}<r_e<400$ kpc (Table \ref{Tab:fit}). Here $r_e$ is the e-folding radius of the exponential model used to fit the surface brightness profile of radio halos \citep[see][]{botteon22}. The latter allows us to compare the radio power of halos and upper limits (Section \ref{sec:bimodality}). The parameters of the correlation agree, within the uncertainties, with those found in \citet{vanweeren21}. However, \citet{vanweeren21} needed to complement the sample with data from the literature to reach a sample of 26 confirmed radio halos. Meanwhile, the number of radio halos in this paper is higher by a factor of almost two and the sample selection is much more homogeneous.      

The correlation shows a relatively large scatter (Table \ref{Tab:fit}). This may be caused by a superposition of clusters undergoing different types of mergers (major or minor mergers), or by clusters that are in different stages of the merger and radio halos with different spectra and sizes. In the context of turbulent re-acceleration models, the scatter of the correlation is expected to increase at low frequency \citep{cassano10b}, owing to the larger probability of forming radio halos and to the larger fraction of USSRHs mixed with classical radio halos. In line with this prediction, we find a hint of increase of the scatter of the correlation at low frequencies (Section \ref{sec:comparison}).
Similar to the work by \citet{cuciti21b} at higher frequencies, we investigated the relation between the scatter of the correlation and the dynamical state of clusters hosting radio halos. To this end, we used the X-ray morphological disturbance, which is a parameter that combines the concentration parameter and the centroid shift, commonly used to infer the dynamics of galaxy clusters. We found a clear trend between the distance of radio halos from the correlation and the X-ray morphological disturbance (Fig.~\ref{fig:scatter_PM}, right). This indicates that the scatter of the correlation can be, at least in part, attributed to the different merging histories of clusters. This could be either due to the increase of the magnetic field strength or of the acceleration efficiency in the most disturbed clusters. The future study of the distribution of radio halo spectra on the mass -- radio power diagram will allow us to discriminate these possibilities. In this respect, the ongoing uGMRT follow up (PI R. Cassano) of the radio halos belonging to this sample will be crucial to measure their spectral properties. Although the X-ray information on clusters without radio halos is incomplete, the vast majority of the upper limits are among the less disturbed clusters of the sample. In order to confirm this, new X-ray observations of clusters without radio halos are required. In this respect, we have obtained (C-priority) XMM-Newton observations of the clusters without diffuse emission. 

In Section \ref{sec:bimodality} we applied two statistical approaches to compare the distribution of clusters with and without radio halos in the $M_{500}-P_{\rm 150\,MHz}$ diagram.
We used the AIC test to compare two models to fit the radio power. In the first model, both radio halos and upper limits were fitted with the same scaling relation, while in the second one radio halos and upper limits were fitted separately. We found that the latter is the one that better describes the distribution of clusters in the mass--radio power diagram.
As a further test, we focused on the distance of clusters from the correlation and implemented a Monte-Carlo test to exclude the possibility that the observed bi-modal distribution of radio halos and upper limits shown in Fig.~\ref{fig:histo} is compatible with clusters being uniformly distributed below the correlation. Our test suggests a very low probability that the observed distributions are simply the result of sensitivity limitations applied to a uniform distribution of clusters in the mass--radio power diagram. This needs to be added to the evidence that for high-mass clusters this hypothesis can be firmly excluded because all the clusters host radio halos (Fig.~\ref{fig:M_z_compl50}, right). Our data support the idea of a clear difference in the density of objects from the correlation to the radio quiet (upper limits) region of the mass--radio power plane. These densities are connected to the amount of time that clusters spend in the different regions of the diagram during their lifetime \citep[see][]{donnert13}. Hence our findings will represent an important input for future simulations of particle acceleration in galaxy clusters, that will have to reproduce the journey of clusters through the $M_{500}-P_{\rm 150\,MHz}$ diagram, matching the existence of a correlation, densely populated with  radio halo clusters and a much less dense region below the correlation populated by clusters without diffuse emission.

In Fig.~\ref{fig:GMRT} (left) we highlighted the radio halos, among those used in this paper, that were discovered by LOFAR \citep{botteon22}. 20, out of 48 radio halos, are new discoveries. 10 of them are in low-mass clusters ($M_{500}<5\times 10^{14}M_\odot$) and this substantially increased the mass range that can be explored with respect to previous works.
In this respect, in Section \ref{sec:comparison} we discussed our statistical findings in comparison with our previous study at higher frequencies \citep{cuciti21b}. In spite of the lower statistics available at 1.4 GHz, we showed that the correlation found at 150 MHz is in agreement with that derived at 1.4 GHz extrapolated to low frequencies assuming a spectral index $\alpha=-1.3$ (Fig.~\ref{fig:GMRT}, right). The number of upper limits, as well as their mass range, has also substantially increased with respect to the past (Fig.~\ref{fig:GMRT}, right). The upper limits obtained with LOFAR are generally deeper than those obtained with the GMRT and this allowed us for the first time to go below $M_{500}\sim 6\times10^{14}M_\odot$ and infer the presence of the correlation down to masses of the order of $M_{500}\sim 3\times10^{14}M_\odot$. 

With the completion of LoTSS we will perform a similar analysis on a sample of clusters 2-3 times larger than the present one, increasing the statistics of radio halos especially at low mass. This will allow us to better constrain the correlation, especially at the low mass end, and the separation between radio halos and upper limits. In fact, just by extending the Montecarlo test presented in Section \ref{sec:MC} to the expected number of clusters with radio halos (178) and upper limits (277) in the complete LoTSS (obtained by rescaling the number of radio halos and upper limits in the current sample by the sky coverage), we find that the probability of clusters of being drawn from a uniform distribution below the correlation is $<0.1$\% even with cut$=-0.5$. This number should be taken just as an indication of the effect of the sample statistics, because its proper assessment needs the correlation and its scatter to be re-evaluated once the survey will be completed. 

\begin{acknowledgements}

VC and GDG acknowledge support from the Alexander von Humboldt Foundation. RJvW acknowledges support from the ERC Starting Grant ClusterWeb 804208. ABotteon acknowledges support from the ERC-StG DRANOEL n. 714245. ABonafede acknowledges support from ERC Stg DRANOEL n. 714245 and MIUR FARE grant ``SMS''. AS is supported by the Women In Science Excel (WISE) programme of the Netherlands Organisation for Scientific Research (NWO), and acknowledges the Kavli IPMU for the continued hospitality. SRON Netherlands Institute for Space Research is supported financially by NWO. RC, GB, FG and MR acknowledge support from INAF mainstream project ‘Galaxy Clusters Science with LOFAR’ 1.05.01.86.05. MB acknowledges support from the Deutsche Forschungsgemeinschaft under Germany's Excellence Strategy - EXC 2121 "Quantum Universe" - 390833306. XZ acknowledges financial support from the ERC Consolidator Grant DarkQuest No. 101002585. MS acknowledges financial contribution from contract ASI-INAF n.2017-14-H.0. and from contract INAF mainstream project 1.05.01.86.10.

LOFAR \citep{vanhaarlem13} is the LOw Frequency ARray designed and constructed by ASTRON. It has observing, data processing, and data storage facilities in several countries, which are owned by various parties (each with their own funding sources), and are collectively operated by the ILT foundation under a joint scientific policy. The ILT resources have benefitted from the following recent major funding sources: CNRS-INSU, Observatoire de Paris and Universit\'{e} d'Orl\'{e}ans, France; BMBF, MIWF-NRW, MPG, Germany; Science Foundation Ireland (SFI), Department of Business, Enterprise and Innovation (DBEI), Ireland; NWO, The Netherlands; The Science and Technology Facilities Council, UK; Ministry of Science and Higher Education, Poland; Istituto Nazionale di Astrofisica (INAF), Italy. This research made use of the Dutch national e-infrastructure with support of the SURF Cooperative (e-infra 180169) and the LOFAR e-infra group, and of the LOFAR-IT computing infrastructure supported and operated by INAF, and by the Physics Dept.~of Turin University (under the agreement with Consorzio Interuniversitario per la Fisica Spaziale) at the C3S Supercomputing Centre, Italy. The J\"{u}lich LOFAR Long Term Archive and the German LOFAR network are both coordinated and operated by the J\"{u}lich Supercomputing Centre (JSC), and computing resources on the supercomputer JUWELS at JSC were provided by the Gauss Centre for Supercomputing e.V. (grant CHTB00) through the John von Neumann Institute for Computing (NIC). This research made use of the University of Hertfordshire high-performance computing facility and the LOFAR-UK computing facility located at the University of Hertfordshire and supported by STFC [ST/P000096/1]. The scientific results reported in this article are based in part on data obtained from the \chandra\ Data Archive. SRON Netherlands Institute for Space Research is supported financially by the Netherlands Organisation for Scientific Research (NWO). 
\end{acknowledgements}

\bibliographystyle{aa}
\bibliography{biblio_virgi.bib}

\begin{thebibliography}{48}
\expandafter\ifx\csname natexlab\endcsname\relax\def\natexlab#1{#1}\fi

\bibitem[{{Akaike}(1974)}]{akaike74}
{Akaike}, H. 1974, IEEE Transactions on Automatic Control, 19, 716

\bibitem[{{Akritas} \& {Bershady}(1996)}]{akritas96}
{Akritas}, M.~G. \& {Bershady}, M.~A. 1996, \apj, 470, 706

\bibitem[{{Basu}(2012)}]{basu12}
{Basu}, K. 2012, \mnras, 421, L112

\bibitem[{{B{\"o}hringer} {et~al.}(2010){B{\"o}hringer}, {Pratt}, {Arnaud},
  {Borgani}, {Croston}, {Ponman}, {Ameglio}, {Temple}, \&
  {Dolag}}]{bohringer10}
{B{\"o}hringer}, H., {Pratt}, G.~W., {Arnaud}, M., {et~al.} 2010, \aap, 514,
  A32

\bibitem[{{Bonafede} {et~al.}(2022){Bonafede}, {Brunetti}, {Rudnick}, {Vazza},
  {Bourdin}, {Giovannini}, {Shimwell}, {Zhang}, {Mazzotta}, {Simionescu},
  {Biava}, {Bonnassieux}, {Brienza}, {Br{\"u}ggen}, {Rajpurohit}, {Riseley},
  {Stuardi}, {Feretti}, {Tasse}, {Botteon}, {Carretti}, {Cassano}, {Cuciti},
  {Gasperin}, {Gastaldello}, {Rossetti}, {Rottgering}, {Venturi}, \&
  {Weeren}}]{bonafede22}
{Bonafede}, A., {Brunetti}, G., {Rudnick}, L., {et~al.} 2022, \apj, 933, 218

\bibitem[{{Bonafede} {et~al.}(2021){Bonafede}, {Brunetti}, {Vazza},
  {Simionescu}, {Giovannini}, {Bonnassieux}, {Shimwell}, {Br{\"u}ggen}, {van
  Weeren}, {Botteon}, {Brienza}, {Cassano}, {Drabent}, {Feretti}, {de
  Gasperin}, {Gastaldello}, {di Gennaro}, {Rossetti}, {Rottgering}, {Stuardi},
  \& {Venturi}}]{bonafede21}
{Bonafede}, A., {Brunetti}, G., {Vazza}, F., {et~al.} 2021, \apj, 907, 32

\bibitem[{{Botteon} {et~al.}(2022){Botteon}, {Shimwell}, {Cassano}, {Cuciti},
  {Zhang}, {Bruno}, {Camillini}, {Natale}, {Jones}, {Gastaldello},
  {Simionescu}, {Rossetti}, {Akamatsu}, {van Weeren}, {Brunetti},
  {Br{\"u}ggen}, {Groeneveld}, {Hoang}, {Hardcastle}, {Ignesti}, {Di Gennaro},
  {Bonafede}, {Drabent}, {R{\"o}ttgering}, {Hoeft}, \& {de
  Gasperin}}]{botteon22}
{Botteon}, A., {Shimwell}, T.~W., {Cassano}, R., {et~al.} 2022, \aap, 660, A78

\bibitem[{{Boxelaar} {et~al.}(2021){Boxelaar}, {van Weeren}, \&
  {Botteon}}]{boxelaar21}
{Boxelaar}, J.~M., {van Weeren}, R.~J., \& {Botteon}, A. 2021, Astronomy and
  Computing, 35, 100464

\bibitem[{{Brunetti} \& {Jones}(2014)}]{brunettijones14}
{Brunetti}, G. \& {Jones}, T.~W. 2014, International Journal of Modern Physics
  D, 23, 1430007

\bibitem[{{Brunetti} {et~al.}(2001){Brunetti}, {Setti}, {Feretti}, \&
  {Giovannini}}]{brunetti01}
{Brunetti}, G., {Setti}, G., {Feretti}, L., \& {Giovannini}, G. 2001, \mnras,
  320, 365

\bibitem[{{Bruno} {et~al.}(2023){Bruno}, {Brunetti}, {Botteon}, {Cuciti},
  {Dallacasa}, {Cassano}, {van Weeren}, {Shimwell}, {Taffoni}, {Russo},
  {Bonafede}, {Br{\"u}ggen}, {Hoang}, {Rottgering}, \& {Tasse}}]{bruno23}
{Bruno}, L., {Brunetti}, G., {Botteon}, A., {et~al.} 2023, \aap, 672, A41

\bibitem[{{Campitiello} {et~al.}(2022){Campitiello}, {Ettori}, {Lovisari},
  {Bartalucci}, {Eckert}, {Rasia}, {Rossetti}, {Gastaldello}, {Pratt},
  {Maughan}, {Pointecouteau}, {Sereno}, {Biffi}, {Borgani}, {De Luca}, {De
  Petris}, {Gaspari}, {Ghizzardi}, {Mazzotta}, \& {Molendi}}]{campitiello22}
{Campitiello}, M.~G., {Ettori}, S., {Lovisari}, L., {et~al.} 2022, \aap, 665,
  A117

\bibitem[{{Cassano}(2010)}]{cassano10b}
{Cassano}, R. 2010, \aap, 517, A10

\bibitem[{{Cassano} \& {Brunetti}(2005)}]{cassanobrunetti05}
{Cassano}, R. \& {Brunetti}, G. 2005, \mnras, 357, 1313

\bibitem[{{Cassano} {et~al.}(2016){Cassano}, {Brunetti}, {Giocoli}, \&
  {Ettori}}]{cassano16}
{Cassano}, R., {Brunetti}, G., {Giocoli}, C., \& {Ettori}, S. 2016, \aap, 593,
  A81

\bibitem[{{Cassano} {et~al.}(2012){Cassano}, {Brunetti}, {Norris},
  {R{\"o}ttgering}, {Johnston-Hollitt}, \& {Trasatti}}]{cassano12}
{Cassano}, R., {Brunetti}, G., {Norris}, R.~P., {et~al.} 2012, \aap, 548, A100

\bibitem[{{Cassano} {et~al.}(2010{\natexlab{a}}){Cassano}, {Brunetti},
  {R{\"o}ttgering}, \& {Br{\"u}ggen}}]{cassano10a}
{Cassano}, R., {Brunetti}, G., {R{\"o}ttgering}, H.~J.~A., \& {Br{\"u}ggen}, M.
  2010{\natexlab{a}}, \aap, 509, A68

\bibitem[{{Cassano} {et~al.}(2023){Cassano}, {Cuciti}, {Brunetti}, {Botteon},
  {Rossetti}, {Bruno}, {Simionescu}, {Gastaldello}, {van Weeren},
  {Br{\"u}ggen}, {Dallacasa}, {Zhang}, {Akamatsu}, {Bonafede}, {Di Gennaro},
  {Shimwell}, {de Gasperin}, {R{\"o}ttgering}, \& {Jones}}]{cassano23}
{Cassano}, R., {Cuciti}, V., {Brunetti}, G., {et~al.} 2023, \aap, 672, A43

\bibitem[{{Cassano} {et~al.}(2013){Cassano}, {Ettori}, {Brunetti},
  {Giacintucci}, {Pratt}, {Venturi}, {Kale}, {Dolag}, \&
  {Markevitch}}]{cassano13}
{Cassano}, R., {Ettori}, S., {Brunetti}, G., {et~al.} 2013, \apj, 777, 141

\bibitem[{{Cassano} {et~al.}(2010{\natexlab{b}}){Cassano}, {Ettori},
  {Giacintucci}, {Brunetti}, {Markevitch}, {Venturi}, \& {Gitti}}]{cassano10}
{Cassano}, R., {Ettori}, S., {Giacintucci}, S., {et~al.} 2010{\natexlab{b}},
  \apjl, 721, L82

\bibitem[{{Cuciti} {et~al.}(2021{\natexlab{a}}){Cuciti}, {Cassano}, {Brunetti},
  {Dallacasa}, {de Gasperin}, {Ettori}, {Giacintucci}, {Kale}, {Pratt}, {van
  Weeren}, \& {Venturi}}]{cuciti21b}
{Cuciti}, V., {Cassano}, R., {Brunetti}, G., {et~al.} 2021{\natexlab{a}}, \aap,
  647, A51

\bibitem[{{Cuciti} {et~al.}(2015){Cuciti}, {Cassano}, {Brunetti}, {Dallacasa},
  {Kale}, {Ettori}, \& {Venturi}}]{cuciti15}
{Cuciti}, V., {Cassano}, R., {Brunetti}, G., {et~al.} 2015, \aap, 580, A97

\bibitem[{{Cuciti} {et~al.}(2021{\natexlab{b}}){Cuciti}, {Cassano}, {Brunetti},
  {Dallacasa}, {van Weeren}, {Giacintucci}, {Bonafede}, {de Gasperin},
  {Ettori}, {Kale}, {Pratt}, \& {Venturi}}]{cuciti21a}
{Cuciti}, V., {Cassano}, R., {Brunetti}, G., {et~al.} 2021{\natexlab{b}}, \aap,
  647, A50

\bibitem[{{de Gasperin} {et~al.}(2021){de Gasperin}, {Williams}, {Best},
  {Br{\"u}ggen}, {Brunetti}, {Cuciti}, {Dijkema}, {Hardcastle}, {Norden},
  {Offringa}, {Shimwell}, {van Weeren}, {Bomans}, {Bonafede}, {Botteon},
  {Callingham}, {Cassano}, {Chy{\.z}y}, {Emig}, {Edler}, {Haverkorn}, {Heald},
  {Heesen}, {Iacobelli}, {Intema}, {Kadler}, {Ma{\l}ek}, {Mevius}, {Miley},
  {Mingo}, {Morabito}, {Sabater}, {Morganti}, {Orr{\'u}}, {Pizzo}, {Prandoni},
  {Shulevski}, {Tasse}, {Vaccari}, {Zarka}, \& {R{\"o}ttgering}}]{degasperin21}
{de Gasperin}, F., {Williams}, W.~L., {Best}, P., {et~al.} 2021, \aap, 648,
  A104

\bibitem[{{Donnert} {et~al.}(2013){Donnert}, {Dolag}, {Brunetti}, \&
  {Cassano}}]{donnert13}
{Donnert}, J., {Dolag}, K., {Brunetti}, G., \& {Cassano}, R. 2013, \mnras, 429,
  3564

\bibitem[{{Duchesne} {et~al.}(2021){Duchesne}, {Johnston-Hollitt}, {Offringa},
  {Pratt}, {Zheng}, \& {Dehghan}}]{duchesne21}
{Duchesne}, S.~W., {Johnston-Hollitt}, M., {Offringa}, A.~R., {et~al.} 2021,
  \pasa, 38, e010

\bibitem[{{Feretti} {et~al.}(2012){Feretti}, {Giovannini}, {Govoni}, \&
  {Murgia}}]{feretti12}
{Feretti}, L., {Giovannini}, G., {Govoni}, F., \& {Murgia}, M. 2012, \aapr, 20,
  54

\bibitem[{{George} {et~al.}(2021){George}, {Kale}, \& {Wadadekar}}]{george21}
{George}, L.~T., {Kale}, R., \& {Wadadekar}, Y. 2021, \mnras, 507, 4487

\bibitem[{{Ghirardini} {et~al.}(2022){Ghirardini}, {Bahar}, {Bulbul}, {Liu},
  {Clerc}, {Pacaud}, {Comparat}, {Liu}, {Ramos Ceja}, {Hoang}, {Ider-Chitham},
  {Klein}, {Merloni}, {Nandra}, {Ota}, {Predehl}, {Reiprich}, {Sanders}, \&
  {Schrabback}}]{ghirardini22}
{Ghirardini}, V., {Bahar}, Y.~E., {Bulbul}, E., {et~al.} 2022, in AAS/High
  Energy Astrophysics Division, Vol.~54, AAS/High Energy Astrophysics Division,
  107.26

\bibitem[{{Jones} {et~al.}(2023){Jones}, {de Gasperin}, {Cuciti}, {Botteon},
  {Zhang}, {Gastaldello}, {Shimwell}, {Simionescu}, {Rossetti}, {Cassano},
  {Akamatsu}, {Bonafede}, {Br{\"u}ggen}, {Brunetti}, {Camillini}, {Di Gennaro},
  {Drabent}, {Hoang}, {Rajpurohit}, {Natale}, {Tasse}, \& {van
  Weeren}}]{jones23}
{Jones}, A., {de Gasperin}, F., {Cuciti}, V., {et~al.} 2023, arXiv e-prints,
  arXiv:2301.07814

\bibitem[{{Kale} {et~al.}(2015){Kale}, {Venturi}, {Giacintucci}, {Dallacasa},
  {Cassano}, {Brunetti}, {Cuciti}, {Macario}, \& {Athreya}}]{kale15}
{Kale}, R., {Venturi}, T., {Giacintucci}, S., {et~al.} 2015, \aap, 579, A92

\bibitem[{{Kale} {et~al.}(2013){Kale}, {Venturi}, {Giacintucci}, {Dallacasa},
  {Cassano}, {Brunetti}, {Macario}, \& {Athreya}}]{kale13}
{Kale}, R., {Venturi}, T., {Giacintucci}, S., {et~al.} 2013, \aap, 557, A99

\bibitem[{{Lima} \& {Hu}(2005)}]{lima05}
{Lima}, M. \& {Hu}, W. 2005, \prd, 72, 043006

\bibitem[{{Lovisari} {et~al.}(2017){Lovisari}, {Forman}, {Jones}, {Ettori},
  {Andrade-Santos}, {Arnaud}, {D{\'e}mocl{\`e}s}, {Pratt}, {Randall}, \&
  {Kraft}}]{lovisari17}
{Lovisari}, L., {Forman}, W.~R., {Jones}, C., {et~al.} 2017, \apj, 846, 51

\bibitem[{{Murgia} {et~al.}(2009){Murgia}, {Govoni}, {Markevitch}, {Feretti},
  {Giovannini}, {Taylor}, \& {Carretti}}]{murgia09}
{Murgia}, M., {Govoni}, F., {Markevitch}, M., {et~al.} 2009, \aap, 499, 679

\bibitem[{{Planck Collaboration} {et~al.}(2016){Planck Collaboration}, {Ade},
  {Aghanim}, {Arnaud}, {Ashdown}, {Aumont}, {Baccigalupi}, {Banday},
  {Barreiro}, {Barrena}, {Bartlett}, {Bartolo}, {Battaner}, {Battye},
  {Benabed}, {Beno{\^\i}t}, {Benoit-L{\'e}vy}, {Bernard}, {Bersanelli},
  {Bielewicz}, {Bikmaev}, {B{\"o}hringer}, {Bonaldi}, {Bonavera}, {Bond},
  {Borrill}, {Bouchet}, {Bucher}, {Burenin}, {Burigana}, {Butler}, {Calabrese},
  {Cardoso}, {Carvalho}, {Catalano}, {Challinor}, {Chamballu}, {Chary},
  {Chiang}, {Chon}, {Christensen}, {Clements}, {Colombi}, {Colombo}, {Combet},
  {Comis}, {Couchot}, {Coulais}, {Crill}, {Curto}, {Cuttaia}, {Dahle},
  {Danese}, {Davies}, {Davis}, {de Bernardis}, {de Rosa}, {de Zotti},
  {Delabrouille}, {D{\'e}sert}, {Dickinson}, {Diego}, {Dolag}, {Dole},
  {Donzelli}, {Dor{\'e}}, {Douspis}, {Ducout}, {Dupac}, {Efstathiou},
  {Eisenhardt}, {Elsner}, {En{\ss}lin}, {Eriksen}, {Falgarone}, {Fergusson},
  {Feroz}, {Ferragamo}, {Finelli}, {Forni}, {Frailis}, {Fraisse}, {Franceschi},
  {Frejsel}, {Galeotta}, {Galli}, {Ganga}, {G{\'e}nova-Santos}, {Giard},
  {Giraud-H{\'e}raud}, {Gjerl{\o}w}, {Gonz{\'a}lez-Nuevo}, {G{\'o}rski},
  {Grainge}, {Gratton}, {Gregorio}, {Gruppuso}, {Gudmundsson}, {Hansen},
  {Hanson}, {Harrison}, {Hempel}, {Henrot-Versill{\'e}},
  {Hern{\'a}ndez-Monteagudo}, {Herranz}, {Hildebrandt}, {Hivon}, {Hobson},
  {Holmes}, {Hornstrup}, {Hovest}, {Huffenberger}, {Hurier}, {Jaffe}, {Jaffe},
  {Jin}, {Jones}, {Juvela}, {Keih{\"a}nen}, {Keskitalo}, {Khamitov}, {Kisner},
  {Kneissl}, {Knoche}, {Kunz}, {Kurki-Suonio}, {Lagache}, {Lamarre}, {Lasenby},
  {Lattanzi}, {Lawrence}, {Leonardi}, {Lesgourgues}, {Levrier}, {Liguori},
  {Lilje}, {Linden-V{\o}rnle}, {L{\'o}pez-Caniego}, {Lubin},
  {Mac{\'\i}as-P{\'e}rez}, {Maggio}, {Maino}, {Mak}, {Mandolesi}, {Mangilli},
  {Martin}, {Mart{\'\i}nez-Gonz{\'a}lez}, {Masi}, {Matarrese}, {Mazzotta},
  {McGehee}, {Mei}, {Melchiorri}, {Melin}, {Mendes}, {Mennella}, {Migliaccio},
  {Mitra}, {Miville-Desch{\^e}nes}, {Moneti}, {Montier}, {Morgante},
  {Mortlock}, {Moss}, {Munshi}, {Murphy}, {Naselsky}, {Nastasi}, {Nati},
  {Natoli}, {Netterfield}, {N{\o}rgaard-Nielsen}, {Noviello}, {Novikov},
  {Novikov}, {Olamaie}, {Oxborrow}, {Paci}, {Pagano}, {Pajot}, {Paoletti},
  {Pasian}, {Patanchon}, {Pearson}, {Perdereau}, {Perotto}, {Perrott},
  {Perrotta}, {Pettorino}, {Piacentini}, {Piat}, {Pierpaoli}, {Pietrobon},
  {Plaszczynski}, {Pointecouteau}, {Polenta}, {Pratt}, {Pr{\'e}zeau}, {Prunet},
  {Puget}, {Rachen}, {Reach}, {Rebolo}, {Reinecke}, {Remazeilles}, {Renault},
  {Renzi}, {Ristorcelli}, {Rocha}, {Rosset}, {Rossetti}, {Roudier}, {Rozo},
  {Rubi{\~n}o-Mart{\'\i}n}, {Rumsey}, {Rusholme}, {Rykoff}, {Sandri}, {Santos},
  {Saunders}, {Savelainen}, {Savini}, {Schammel}, {Scott}, {Seiffert},
  {Shellard}, {Shimwell}, {Spencer}, {Stanford}, {Stern}, {Stolyarov},
  {Stompor}, {Streblyanska}, {Sudiwala}, {Sunyaev}, {Sutton}, {Suur-Uski},
  {Sygnet}, {Tauber}, {Terenzi}, {Toffolatti}, {Tomasi}, {Tramonte},
  {Tristram}, {Tucci}, {Tuovinen}, {Umana}, {Valenziano}, {Valiviita}, {Van
  Tent}, {Vielva}, {Villa}, {Wade}, {Wandelt}, {Wehus}, {White}, {Wright},
  {Yvon}, {Zacchei}, \& {Zonca}}]{planck16}
{Planck Collaboration}, {Ade}, P.~A.~R., {Aghanim}, N., {et~al.} 2016, \aap,
  594, A27

\bibitem[{{Rasia} {et~al.}(2013){Rasia}, {Meneghetti}, \& {Ettori}}]{rasia13}
{Rasia}, E., {Meneghetti}, M., \& {Ettori}, S. 2013, The Astronomical Review,
  8, 40

\bibitem[{{Santos} {et~al.}(2008){Santos}, {Rosati}, {Tozzi}, {B{\"o}hringer},
  {Ettori}, \& {Bignamini}}]{santos08}
{Santos}, J.~S., {Rosati}, P., {Tozzi}, P., {et~al.} 2008, \aap, 483, 35

\bibitem[{{Sereno}(2016)}]{sereno16}
{Sereno}, M. 2016, \mnras, 455, 2149

\bibitem[{{Sereno} \& {Ettori}(2015)}]{sereno15}
{Sereno}, M. \& {Ettori}, S. 2015, \mnras, 450, 3675, (CoMaLit-IV)

\bibitem[{{Shimwell} {et~al.}(2017){Shimwell}, {R{\"o}ttgering}, {Best},
  {Williams}, {Dijkema}, {de Gasperin}, {Hardcastle}, {Heald}, {Hoang},
  {Horneffer}, {Intema}, {Mahony}, {Mandal}, {Mechev}, {Morabito}, {Oonk},
  {Rafferty}, {Retana-Montenegro}, {Sabater}, {Tasse}, {van Weeren},
  {Br{\"u}ggen}, {Brunetti}, {Chy{\.z}y}, {Conway}, {Haverkorn}, {Jackson},
  {Jarvis}, {McKean}, {Miley}, {Morganti}, {White}, {Wise}, {van Bemmel},
  {Beck}, {Brienza}, {Bonafede}, {Calistro Rivera}, {Cassano}, {Clarke},
  {Cseh}, {Deller}, {Drabent}, {van Driel}, {Engels}, {Falcke}, {Ferrari},
  {Fr{\"o}hlich}, {Garrett}, {Harwood}, {Heesen}, {Hoeft}, {Horellou},
  {Israel}, {Kapi{\'n}ska}, {Kunert-Bajraszewska}, {McKay}, {Mohan},
  {Orr{\'u}}, {Pizzo}, {Prandoni}, {Schwarz}, {Shulevski}, {Sipior}, {Smith},
  {Sridhar}, {Steinmetz}, {Stroe}, {Varenius}, {van der Werf}, {Zensus}, \&
  {Zwart}}]{shimwell17}
{Shimwell}, T.~W., {R{\"o}ttgering}, H.~J.~A., {Best}, P.~N., {et~al.} 2017,
  \aap, 598, A104

\bibitem[{{Shimwell} {et~al.}(2019){Shimwell}, {Tasse}, {Hardcastle}, {Mechev},
  {Williams}, {Best}, {R{\"o}ttgering}, {Callingham}, {Dijkema}, {de Gasperin},
  {Hoang}, {Hugo}, {Mirmont}, {Oonk}, {Prandoni}, {Rafferty}, {Sabater},
  {Smirnov}, {van Weeren}, {White}, {Atemkeng}, {Bester}, {Bonnassieux},
  {Br{\"u}ggen}, {Brunetti}, {Chy{\.z}y}, {Cochrane}, {Conway}, {Croston},
  {Danezi}, {Duncan}, {Haverkorn}, {Heald}, {Iacobelli}, {Intema}, {Jackson},
  {Jamrozy}, {Jarvis}, {Lakhoo}, {Mevius}, {Miley}, {Morabito}, {Morganti},
  {Nisbet}, {Orr{\'u}}, {Perkins}, {Pizzo}, {Schrijvers}, {Smith}, {Vermeulen},
  {Wise}, {Alegre}, {Bacon}, {van Bemmel}, {Beswick}, {Bonafede}, {Botteon},
  {Bourke}, {Brienza}, {Calistro Rivera}, {Cassano}, {Clarke}, {Conselice},
  {Dettmar}, {Drabent}, {Dumba}, {Emig}, {En{\ss}lin}, {Ferrari}, {Garrett},
  {G{\'e}nova-Santos}, {Goyal}, {G{\"u}rkan}, {Hale}, {Harwood}, {Heesen},
  {Hoeft}, {Horellou}, {Jackson}, {Kokotanekov}, {Kondapally},
  {Kunert-Bajraszewska}, {Mahatma}, {Mahony}, {Mandal}, {McKean}, {Merloni},
  {Mingo}, {Miskolczi}, {Mooney}, {Nikiel-Wroczy{\'n}ski}, {O'Sullivan},
  {Quinn}, {Reich}, {Roskowi{\'n}ski}, {Rowlinson}, {Savini}, {Saxena},
  {Schwarz}, {Shulevski}, {Sridhar}, {Stacey}, {Urquhart}, {van der Wiel},
  {Varenius}, {Webster}, \& {Wilber}}]{shimwell19}
{Shimwell}, T.~W., {Tasse}, C., {Hardcastle}, M.~J., {et~al.} 2019, \aap, 622,
  A1

\bibitem[{{van Haarlem} {et~al.}(2013){van Haarlem}, {Wise}, {Gunst}, {Heald},
  {McKean}, {Hessels}, {de Bruyn}, {Nijboer}, {Swinbank}, {Fallows},
  {Brentjens}, {Nelles}, {Beck}, {Falcke}, {Fender}, {H{\"o}randel},
  {Koopmans}, {Mann}, {Miley}, {R{\"o}ttgering}, {Stappers}, {Wijers},
  {Zaroubi}, {van den Akker}, {Alexov}, {Anderson}, {Anderson}, {van Ardenne},
  {Arts}, {Asgekar}, {Avruch}, {Batejat}, {B{\"a}hren}, {Bell}, {Bell}, {van
  Bemmel}, {Bennema}, {Bentum}, {Bernardi}, {Best}, {B{\^i}rzan}, {Bonafede},
  {Boonstra}, {Braun}, {Bregman}, {Breitling}, {van de Brink}, {Broderick},
  {Broekema}, {Brouw}, {Br{\"u}ggen}, {Butcher}, {van Cappellen}, {Ciardi},
  {Coenen}, {Conway}, {Coolen}, {Corstanje}, {Damstra}, {Davies}, {Deller},
  {Dettmar}, {van Diepen}, {Dijkstra}, {Donker}, {Doorduin}, {Dromer}, {Drost},
  {van Duin}, {Eisl{\"o}ffel}, {van Enst}, {Ferrari}, {Frieswijk}, {Gankema},
  {Garrett}, {de Gasperin}, {Gerbers}, {de Geus}, {Grie{\ss}meier}, {Grit},
  {Gruppen}, {Hamaker}, {Hassall}, {Hoeft}, {Holties}, {Horneffer}, {van der
  Horst}, {van Houwelingen}, {Huijgen}, {Iacobelli}, {Intema}, {Jackson},
  {Jelic}, {de Jong}, {Juette}, {Kant}, {Karastergiou}, {Koers}, {Kollen},
  {Kondratiev}, {Kooistra}, {Koopman}, {Koster}, {Kuniyoshi}, {Kramer},
  {Kuper}, {Lambropoulos}, {Law}, {van Leeuwen}, {Lemaitre}, {Loose}, {Maat},
  {Macario}, {Markoff}, {Masters}, {McFadden}, {McKay-Bukowski}, {Meijering},
  {Meulman}, {Mevius}, {Middelberg}, {Millenaar}, {Miller-Jones}, {Mohan},
  {Mol}, {Morawietz}, {Morganti}, {Mulcahy}, {Mulder}, {Munk}, {Nieuwenhuis},
  {van Nieuwpoort}, {Noordam}, {Norden}, {Noutsos}, {Offringa}, {Olofsson},
  {Omar}, {Orr{\'u}}, {Overeem}, {Paas}, {Pandey-Pommier}, {Pandey}, {Pizzo},
  {Polatidis}, {Rafferty}, {Rawlings}, {Reich}, {de Reijer}, {Reitsma},
  {Renting}, {Riemers}, {Rol}, {Romein}, {Roosjen}, {Ruiter}, {Scaife}, {van
  der Schaaf}, {Scheers}, {Schellart}, {Schoenmakers}, {Schoonderbeek},
  {Serylak}, {Shulevski}, {Sluman}, {Smirnov}, {Sobey}, {Spreeuw}, {Steinmetz},
  {Sterks}, {Stiepel}, {Stuurwold}, {Tagger}, {Tang}, {Tasse}, {Thomas},
  {Thoudam}, {Toribio}, {van der Tol}, {Usov}, {van Veelen}, {van der Veen},
  {ter Veen}, {Verbiest}, {Vermeulen}, {Vermaas}, {Vocks}, {Vogt}, {de Vos},
  {van der Wal}, {van Weeren}, {Weggemans}, {Weltevrede}, {White}, {Wijnholds},
  {Wilhelmsson}, {Wucknitz}, {Yatawatta}, {Zarka}, {Zensus}, \& {van
  Zwieten}}]{vanhaarlem13}
{van Haarlem}, M.~P., {Wise}, M.~W., {Gunst}, A.~W., {et~al.} 2013, \aap, 556,
  A2

\bibitem[{{van Weeren} {et~al.}(2019){van Weeren}, {de Gasperin}, {Akamatsu},
  {Br{\"u}ggen}, {Feretti}, {Kang}, {Stroe}, \& {Zandanel}}]{vanweeren19}
{van Weeren}, R.~J., {de Gasperin}, F., {Akamatsu}, H., {et~al.} 2019, \ssr,
  215, 16

\bibitem[{{van Weeren} {et~al.}(2021){van Weeren}, {Shimwell}, {Botteon},
  {Brunetti}, {Br{\"u}ggen}, {Boxelaar}, {Cassano}, {Di Gennaro},
  {Andrade-Santos}, {Bonnassieux}, {Bonafede}, {Cuciti}, {Dallacasa}, {de
  Gasperin}, {Gastaldello}, {Hardcastle}, {Hoeft}, {Kraft}, {Mandal},
  {Rossetti}, {R{\"o}ttgering}, {Tasse}, \& {Wilber}}]{vanweeren21}
{van Weeren}, R.~J., {Shimwell}, T.~W., {Botteon}, A., {et~al.} 2021, \aap,
  651, A115

\bibitem[{{Venturi} {et~al.}(2007){Venturi}, {Giacintucci}, {Brunetti},
  {Cassano}, {Bardelli}, {Dallacasa}, \& {Setti}}]{venturi07}
{Venturi}, T., {Giacintucci}, S., {Brunetti}, G., {et~al.} 2007, \aap, 463, 937

\bibitem[{{Venturi} {et~al.}(2008){Venturi}, {Giacintucci}, {Dallacasa},
  {Cassano}, {Brunetti}, {Bardelli}, \& {Setti}}]{venturi08}
{Venturi}, T., {Giacintucci}, S., {Dallacasa}, D., {et~al.} 2008, \aap, 484,
  327

\bibitem[{{Zhang} {et~al.}(2023){Zhang}, {Simionescu}, {Gastaldello}, {Eckert},
  {Camillini}, {Natale}, {Rossetti}, {Brunetti}, {Akamatsu}, {Botteon},
  {Cassano}, {Cuciti}, {Bruno}, {Shimwell}, {Jones}, {Kaastra}, {Ettori},
  {Br{\"u}ggen}, {de Gasperin}, {Drabent}, {van Weeren}, \&
  {R{\"o}ttgering}}]{zhang23}
{Zhang}, X., {Simionescu}, A., {Gastaldello}, F., {et~al.} 2023, \aap, 672, A42

\end{thebibliography}

\clearpage
\onecolumn
\begin{appendix}
\clearpage

\section{Morphological parameter M versus disturbance}
\label{app:rasia}
Here we derive the morphological parameter $M_C$ from the combination of $c$ and $w$, using the definition by \citet{campitiello22}:

\begin{equation}
    M_C = \frac{\langle \mathrm{log}c\rangle - \mathrm{log}c}{\sigma_{\mathrm{log}_{10}c}} + \frac{\mathrm{log}w - \langle \mathrm{log}w\rangle}{\sigma_{\mathrm{log}_{10}w}} .
\end{equation}
With this definition, relaxed systems should have low values of $M_C$, while disturbed systems should have high values of $M_C$. 

\begin{figure*}
\centering
\includegraphics[scale=0.4,trim={0cm 0cm 0cm 0cm},clip]{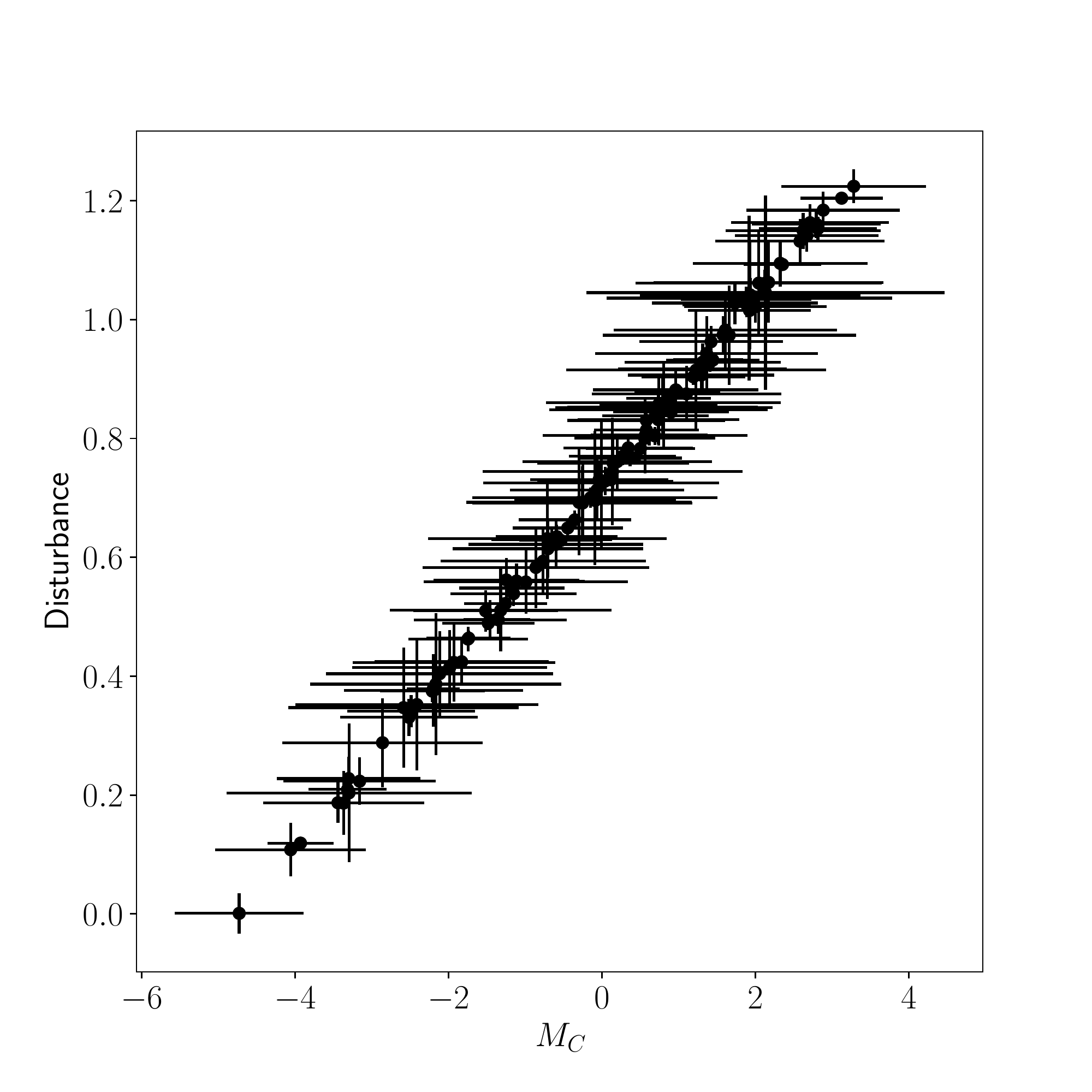}
\includegraphics[scale=0.4,trim={0cm 0cm 0cm 0cm},clip]{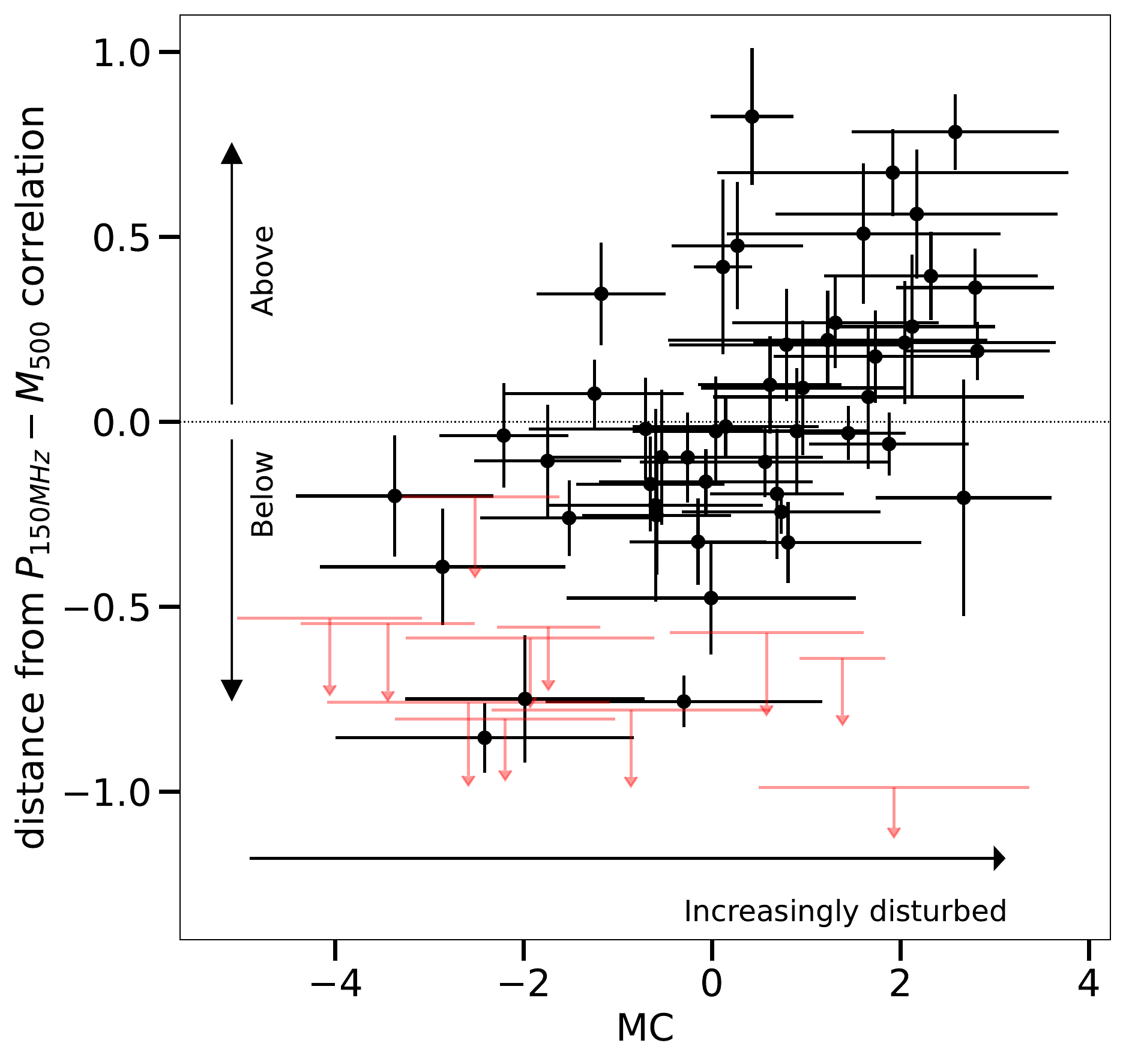}
\caption{Morphological parameter $M$ versus disturbance. Left: Disturbance as derived in Section \ref{Sec:scatter} versus $M$. Right: distance of radio halos from the mass--radio power correlation (BCES Y|X method) vs. M parameter.}
\label{fig:scatter_PM_Rasia}
\end{figure*}

In Fig.~\ref{fig:scatter_PM_Rasia} (left) we show the values of the parameter $M_C$ versus the disturbance, as derived in Section \ref{Sec:scatter}. We found a well-defined correlation which indicates that there is good agreement between the two approaches in classifying the dynamical state of clusters. For completeness, we show in Fig.~\ref{fig:scatter_PM_Rasia} (right) the distance of radio halos and upper limits (those with available X-ray information) versus the $M_C$ parameter. We confirm the presence of the trend discussed in Section \ref{Sec:scatter}, also using this additional parameter.

\section{Upper limits}
\label{app:UL}
Let us consider a variable $Y$ in the domain $Y_\text{min} < Y < Y_\text{max}$.
We assume that $Y$ is a scattered proxy of the variable $X$, the scatter $\sigma_Y$ being approximately normal. We can write the probability as
\begin{equation}
p(X,Y) = \frac{1}{\sqrt{2\pi}\sigma_Y} \frac{1}{Y_\text{max} - Y_\text{min}} \exp \left[ -\frac{1}{2}\left(\frac{Y-X}{\sigma_Y}\right)^2 \right] \;.
\end{equation}
If $X$ is detected whereas $Y$ is not, we can marginalise over $Y$,
\begin{equation}
p(X) =  \frac{1}{2 (Y_\text{max} - Y_\text{min})} \left[ \text{erf}\left( \frac{X- Y_\text{min}}{\sqrt{2}\sigma_Y} \right)  - \text{erf}\left( \frac{X- Y_\text{max}}{\sqrt{2}\sigma_Y} \right) \right] \;.
\end{equation}
The above expression was used in Eq.~(\ref{eq_aic_2}) for the likelihood of the upper limits.

\end{appendix}
\end{document}